\documentclass[twocolumn,graphics,floatfix,a4paper,prl,nofootinbib]{revtex4}
\usepackage{graphicx, epsfig,amsmath, amssymb,mathtools,braket,bbm,turnstile,color,fancyhdr,marvosym}

\newcommand{\Fig}[1]{Fig.~\ref{#1}}

\newcommand{\Eq}[1]{Eq.(\ref{#1})}
\newcommand{\Eqs}[1]{Eqs.(\ref{#1})}

\newcommand{\bit}{\begin{itemize} \setlength{\itemsep}{0ex} \setlength{\topsep}{0ex} }
\newcommand{\eit}{\end{itemize}}
\newcommand{\be}{\begin{equation}}
\newcommand{\ee}{\end{equation}}
\newcommand{\bea}{\begin{eqnarray}}
\newcommand{\eea}{\end{eqnarray}}
\newcommand{\ba}{\begin{align}}
\newcommand{\ea}{\end{align}}
\newcommand{\SKIP}[1]{}
\newcommand{\T}{\textrm{\TTsteel}}
\newcommand{\nx}{\lVert \nabla E \rVert}

\newcommand{\mybar}{}

\def \ra{\rightarrow}

\def \eps{\epsilon}

\def \tr{{\rm tr}}

\begin{document}
\title{Continuous Matrix Product States for Quantum Fields: \\
an Energy Minimization Algorithm}
\author{Martin Ganahl}
\email{martin.ganahl@gmail.com}
\affiliation{Perimeter Institute for Theoretical Physics, 31 Caroline Street North, Waterloo, ON N2L 2Y5, Canada}
\author{Juli\'an Rinc\'on}
\affiliation{Perimeter Institute for Theoretical Physics, 31 Caroline Street North, Waterloo, ON N2L 2Y5, Canada}
\author{Guifre Vidal}
\affiliation{Perimeter Institute for Theoretical Physics, 31 Caroline Street North, Waterloo, ON N2L 2Y5, Canada}
\begin{abstract}
The generalization of matrix product states (MPS) to continuous systems, as proposed in the breakthrough paper [F. Verstraete, J.I. Cirac, Phys. Rev. Lett. 104, 190405(2010)], provides a powerful variational 
ansatz for the ground state of strongly interacting quantum field theories in one spatial dimension. A continuous MPS (cMPS) approximation to the ground state can be obtained by simulating an Euclidean time 
evolution. In this Letter we propose a cMPS optimization algorithm based instead on energy minimization by gradient methods, and demonstrate its performance by applying it to the Lieb Liniger model (an 
integrable model of an interacting bosonic field) directly in the thermodynamic limit. We observe a very significant computational speed-up, of more than two orders of magnitude, with respect to simulating an Euclidean time evolution. As a result, much larger cMPS bond dimension $D$ can be reached (e.g. $D = 256$ with moderate computational resources) thus helping unlock the full potential of the cMPS representation for ground state studies.
\end{abstract}

\maketitle

Over the last 25 years, progress in our understanding of quantum spin chains and other strongly interacting quantum many-body systems in one spatial dimension has been dominated by a variational ansatz: 
the \textit{matrix product state} (MPS) \cite{fannes_finitely_1992,white_density_1992,schollwock_density-matrix_2011,verstraete_matrix_2009}. 
The wave function $\ket{\Psi}$ of a quantum spin chain made of $N$ spin-$1/2$ degrees of freedom depends on $2^N$ complex parameters $\Psi_{i_1\cdots i_N}$,
\begin{equation} \label{eq:Psi}
\ket{\Psi} = \sum_{i_1=0}^1 \sum_{i_2=0}^1 \cdots \sum_{i_N=0}^1 \Psi_{i_1 i_2 \cdots i_N} \ket{i_1 i_2 \cdots i_N}.
\end{equation}
Accordingly, an exact numerical simulation has a computational cost that grows exponentially with the size $N$ of the chain. In an MPS, the $2^N$ 
coefficients are expressed in terms of the trace of a product of matrices. For instance, in a translation invariant system the MPS reads 
\begin{equation} \label{eq:mps}
\Psi_{i_1 i_2 \cdots i_N} = \tr \left[ A^{i_1} A^{i_2} \cdots A^{i_N}\right],
\end{equation}
where $A^{0}$ and $A^{1}$ are $D\times D$ complex matrices. Thus, the state $\ket{\Psi}$ of $N$ spins is specified by just $O(D^2)$ variational parameters, allowing for the study of arbitrarily large, 
even infinite, systems \cite{mcculloch_infinite_2008,vidal_classical_2007}. 

A generic state of the spin chain can not be expressed as an MPS, because the bond dimension $D$ limits how entangled $\ket{\Psi}$ can be. However, ground states of local Hamiltonians happen to be weakly entangled (e.g. they obey an entanglement area law \cite{hastings_area_2007,eisert_textitcolloquium_2010})
in a way that allows for an accurate approximation by an MPS. Given a Hamiltonian $H$, White's revolutionary 
\textit{density matrix renormalization group} (DMRG) \cite{white_density_1992,white_density-matrix_1993}
algorithm provided the first systematic way of obtaining a ground state MPS approximation by minimizing the energy, see also \cite{white_density-matrix_1993}. 
Subsequently, Refs. \cite{vidal_efficient_2004,vidal_efficient_2003} proposed an algorithm to simulate time evolution with an 
MPS, which in Euclidean time also produces a ground state approximation, see also Refs. \cite{daley_time-dependent_2004,white_real-time_2004,feiguin_time-step_2005}. 
An improved formulation of the time evolution simulation by MPS was obtained in terms of the
 \textit{time-dependent variational principle} (TDVP) \cite{haegeman_time-dependent_2011}.

The continuous version of an MPS (cMPS), introduced by Verstraete and Cirac \cite{verstraete_continuous_2010,haegeman_calculus_2013}, 
has the potential of duplicating, in the context of quantum field theories in the continuum, the enormous success of the MPS on the lattice. A cMPS expresses 
the wave function $\ket{\Psi}$ of a quantum field on a circle of radius $L$ as a path ordered exponential $\mathcal{P}e$ of 
the fields that define the theory. 
For a bosonic, translation invariant system it reads
\begin{equation}
  \ket{\Psi}=\tr \left[\mathcal{P}e^{\int_{0}^{L}dx\, Q\otimes \mathbbm{1}+R\otimes\psi^{\dagger}(x)}\right] \ket{\Omega},
  \label{eq:cmps}
\end{equation}
where $\psi^{\dagger}(x)$ is the bosonic field creation operator, 
\begin{equation}
\left[ \psi(x), \psi(y)\right] = 0,~~~~\left[ \psi(x), \psi(y)^{\dagger}\right] = \delta(x-y),
\end{equation}
$\ket{\Omega}$ is the empty state, \textit{i.e.} $\psi(x)\ket{\Omega} = 0$, and $Q$ and $R$ are $D\times D$ complex matrices. 
Again, the wave function $\ket{\Psi}$ is parameterized by just $O(D^2)$ parameters. A cMPS approximation to the ground state of a continuum Hamiltonian $H$ can then be obtained by simulating an Euclidean time evolution with TDVP adapted to cMPS
\cite{haegeman_time-dependent_2011}. While this algorithm and its variations work reasonably well for small $D$ up to $D \sim 50$
\cite{haegeman_applying_2010,draxler_particles_2013,rincon_lieb-liniger_2015,draxler_atomtronics_2016}, 
their performance is poor compared to lattice MPS techniques. 

In this Letter we propose an energy minimization algorithm to find a cMPS approximation for ground states, based on gradient descent techniques, and demonstrate its performance with the Lieb Liniger model in the thermodynamic limit ($L \rightarrow \infty$). 
We also propose a useful cMPS initialization scheme, of interest on its own, 
based on lattice MPS algorithms. 
These proposals result in a very significant computational speed-up with respect to Euclidean time evolution -- e.g. converging 
a cMPS with bond dimension $D=256$ requires less time than a $D= 64$ computation with TDVP. For simplicity we consider a single bosonic field. Generalization to a fermionic field and 
to multiple fields is straightforward.

\textit{Continuum limit and central canonical form}.--- In order to describe the algorithm, we must first adjust the notation in two ways. Firstly, following \cite{verstraete_continuous_2010}, 
we discretize the interval $[0,L)$ in \Eq{eq:cmps} into a regular lattice made of $N \equiv L /\epsilon$ sites and with inter-site spacing $\epsilon$, 
and produce an MPS with matrices $A^0$ and $A^1$ \cite{moreA} given, in vectorized form, by 
\begin{eqnarray}\label{eq:disccmps}
\left( \begin{array}{c} A^{0} \\ A^{1} \end{array} \right) = 
\left( \begin{array}{c} \mathbbm{1}+\eps Q \\ \sqrt{\eps} R \end{array} \right), 
\end{eqnarray}
such that the original cMPS is recovered in the limit 
$\epsilon \rightarrow 0$ \cite{verstraete_continuous_2010}. Here, $A^0$ and $A^1$ corresponds to having 0 or 1 particle at the 
lattice site.
This lattice visualization is useful in order to manipulate the cMPS with regular MPS techniques, provided the latter have a well-defined continuum limit ($\epsilon \rightarrow 0$). Secondly, we use the lattice visualization to re-express the cMPS of an infinite system ($L \rightarrow \infty$) in the \textit{central canonical form} \cite{stoudenmire_real-space_2013}, \Eq{eq:centralmps} below. For this purpose, we consider the Schmidt decomposition of $\ket{\Psi}$ according to a left/right partition of the resulting infinite lattice \cite{InfiniteChain},
\begin{equation} \label{eq:schmidt}
\ket{\Psi} = \sum_{\alpha=1}^D \lambda_{\alpha} \ket{\Psi_{l,\alpha}}\ket{\Psi_{r,\alpha}},~~~~~\lambda_{1} \geq \cdots \geq \lambda_D > 0,
\end{equation}
 and denote by $\lambda$ a diagonal matrix with the $D$ Schmidt coefficients $\{\lambda_1, \cdots, \lambda_D\}$ in its diagonal. 
In the central canonical form, the MPS $\ket{\Psi}$ is expressed as the infinite product of (vectorized) matrices
\begin{equation} \label{eq:centralmps}
\ket{\Psi} \sim \cdots \lambda^{-1}\!
\left( \begin{array}{c} A_c^{0} \\ A_c^{1} \end{array} \right) \lambda^{-1} \!
\left( \begin{array}{c} A_c^{0} \\ A_c^{1} \end{array} \right) \lambda^{-1}\!
\left( \begin{array}{c} A_c^{0} \\ A_c^{1} \end{array} \right) \lambda^{-1}\!\cdots\,.
\end{equation}
The matrices $A_c^0$ and $A_c^1$ are chosen such that
\begin{align} \label{eq:translate}
  \left( \begin{array}{c} A^{0} \\ A^{1} \end{array} \right) \equiv
  \left( \begin{array}{c} A_c^{0} \\ A_c^{1} \end{array} \right) \lambda^{-1} \; \mbox{and}\;
  \left( \begin{array}{c} B^{0} \\ B^{1} \end{array} \right) \equiv
  \lambda^{-1}\left( \begin{array}{c} A_c^{0} \\ A_c^{1} \end{array} \right)
\end{align}
are in the \textit{left} and \textit{right canonical form} \cite{schollwock_density-matrix_2011}, namely 
\begin{eqnarray} 
 \left(A^{0}\right)^{\dagger}A^{0} + \left(A^{1}\right)^{\dagger}A^{1} &=& \mathbbm{1},\\
  B^{0} (B^{0})^{\dagger} + B^{1}(B^{1})^{\dagger} &=& \mathbbm{1}.
\end{eqnarray}
From \Eqs{eq:centralmps}-(\ref{eq:translate}) the standard MPS form \Eq{eq:mps} (in the $L\rightarrow \infty$ limit) for e.g. a {\it left normalized} MPS is recovered,
\begin{equation}
\ket{\Psi} \sim \cdots 
\left( \begin{array}{c} A^{0} \\ A^{1} \end{array} \right) 
\left( \begin{array}{c} A^{0} \\ A^{1} \end{array} \right) 
\left( \begin{array}{c} A^{0} \\ A^{1} \end{array} \right) \cdots.
\end{equation}

In the central canonical form, familiar to DMRG and MPS practitioners working with so-called single-site updates, a change in the matrices $A_c^0$ and $A_c^1$ on a single site produces an equivalent change 
in $\ket{\Psi}$, in the sense that the scalar product in the lattice Hilbert space and in the effective one-site Hilbert space are equivalent (they are related by an isometry). This is important when applying gradient methods, because
two gradients, calculated in two different gauges of the same state, are in general not related by a gauge transformation and are not equivalent. 
The importance of the central gauge has been realized early on in DMRG \cite{white_density_1992} and also time evolution 
methods \cite{vidal_efficient_2003,vidal_classical_2007,stoudenmire_real-space_2013,haegeman_unifying_2016,vanderstraeten_tangent,halimeh_enriching_2016}.

Finally, in the continuum limit, the central canonical form is given by (\textit{c.f.} \Eqs{eq:disccmps} and (\ref{eq:translate}))
\begin{eqnarray}\label{eq:centralcmps}
\left( \begin{array}{c} A_c^{0} \\ A_c^{1} \end{array} \right) = 
\left( \begin{array}{c} \lambda+\eps Q_c \\ \sqrt{\eps} R_c \end{array} \right). 
\end{eqnarray}

\textit{Gradient descent}.--- Given a quantum field Hamiltonian $H$, see e.g. \Eq{eq:Ham}, our goal is to iteratively optimize the cMPS in such a way that the energy
\begin{equation}\label{eq:E}
 E(\lambda,Q_c,R_c) \equiv \frac{\bra{\Psi}H\ket{\Psi}}{\braket{\Psi|\Psi}}
\end{equation} 
is minimized. Each iteration updates a triplet $(\lambda^{[n]}, Q^{[n]}_c, R^{[n]}_c)$ and is made of two steps. \textit{(i)} First, keeping $\lambda$ fixed, we update $Q_c$ and $R_c$ in the direction of \textit{steepest descent} given by the gradient, namely 
\begin{eqnarray} \label{eq:QR}
\left( \begin{array}{c} \tilde{Q}^{[n]} \\ \\ \tilde{R}^{[n]}\end{array} \right) = 
\left( \begin{array}{c} Q^{[n]}_c \\ \\ R^{[n]}_c\end{array} \right) - \alpha_n 
\left( \begin{array}{c} \partial E/\partial Q_c^* \\ \\ \partial E/\partial R_c^* \end{array} \right),
\end{eqnarray}
where $\alpha_n>0$ is some adjustable parameter and ${}^*$ denotes complex conjugation. Crucially, the gradients $\partial E / \partial Q^*_c$ and $\partial E / \partial R^*_c$ can be efficiently computed 
using standard cMPS contraction techniques. We dynamically choose the largest possible factor $\alpha_n$ by requiring consistency with some simple stability conditions (alternatively, 
$\alpha_n$ can be determined by a line search). \textit{(ii)} Then, from $(\lambda^{[n]},\tilde{Q}^{[n]},\tilde{R}^{[n]})$ we obtain $(\lambda^{[n+1]},Q_c^{[n+1]},R_c^{[n+1]})$ by bringing the cMPS representation back into the 
central canonical form. This completes an iteration, which has a cost comparable to one time step 
in TDVP. We emphasize that all manipulations are implemented directly in the continuum limit 
i.e. $\eps$ is treated as an analytic parameter throughout the optimization, and the $\eps\ra 0$ limit can be taken {\it exactly} 
due to exact cancellation of all divergencies.

Overall, the proposed energy minimization algorithm proceeds as follows (see \cite{suppl} for technical details). 

\textit{(a) Initialization:} An initial triplet of matrices $(\lambda^{[0]},Q_c^{[0]},R_c^{[0]})$ is obtained, either from a random initialization or, as in this Letter, through \Eq{eq:centralcmps} from an MPS optimized on the lattice. 

\textit{(b) Iteration:} The above update $(\lambda ^{[n]},Q_c^{[n]},R_c^{[n]}) \mapsto (\lambda^{[n+1]},Q_c^{[n+1]},R_c^{[n+1]})$ is iteratively applied until attaining a suitably converged triplet $(\lambda, Q_c,R_c)$. 

\textit{(c) Final output:} A standard cMPS representation as in \Eq{eq:cmps} is recovered by transforming the result into $(Q,R)$. For instance, $(Q,R) = (Q_c\lambda^{-1},R_c\lambda^{-1})$ as in \Eq{eq:translate} for a final cMPS in the left canonical form (see
also \cite{suppl}).

As usual in such optimization methods, convergence can be accelerated by replacing the gradient descent in \Eq{eq:QR} with e.g. a non-linear conjugate gradient update, 
which re-uses the gradient computed in previous steps (see \cite{milsted_matrix_2013,suppl}).

\textit{Example.}--- To benchmark the above algorithm, we have applied it to obtain a cMPS approximation to the ground state of the Lieb Liniger model \cite{lieb_exact_1963,lieb_exact_1963-1},
\begin{align}
  H &= \int dx\, \Big(\frac{1}{2m} \partial_x\psi^\dagger(x) \partial_x\psi(x) + \mu   \psi^\dagger(x) \psi(x)\nonumber\\
  &+ g \,\psi^\dagger(x) \psi^\dagger(x) \psi(x) \psi(x)\Big),
\label{eq:Ham}
\end{align}
which is both of theoretical and of experimental interest and has been realized in several cold atom experiments
\cite{jaksch_cold_1998,bloch_ultracold_2005,kinoshita_quantum_2006,cazalilla_one_2011,haller_pinning_2010,meinert_bloch_2016}. 
This integrable Hamiltonian has a critical, gapless ground state that can be described by Luttinger liquid theory \cite{cazalilla_one_2011} 
and can be exactly solved by Bethe ansatz \cite{lieb_exact_1963,lieb_exact_1963-1,prolhac_ground_2016,lang_ground-state_2016,ristivojevic_excitation_2014,zvonarev_bethe}. 

\begin{figure}
  \includegraphics[width=1\columnwidth]{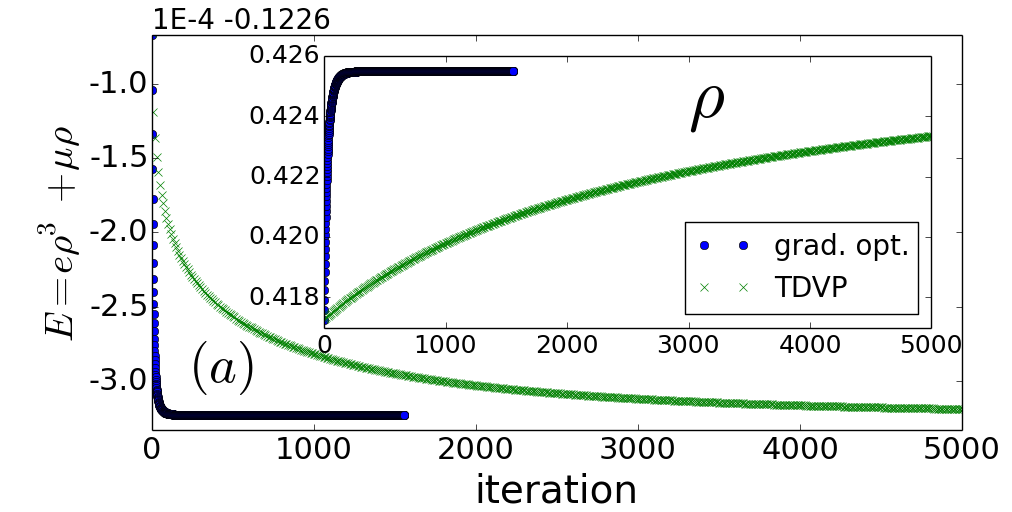}
  \includegraphics[width=1\columnwidth]{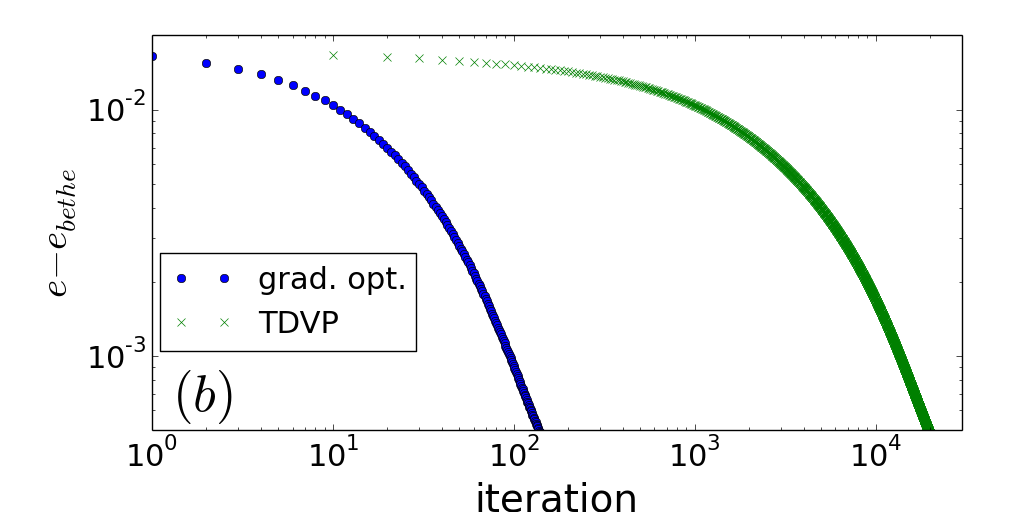}
  \caption{
    Convergence of gradient optimization and of TDVP, for $D=16$ and $(\mu, g, m) = (-0.5, 1.0, 0.5)$. We used $d\tau=0.01$ as time step for TDVP and $\alpha=0.01$ for the steepest descent 
    optimization \cite{suppl}. The time per iteration for either method is ~0.2s.
    (a) Energy density $E$ (main figure) and particle density $\rho$ (inset) as a function of iteration number. 
    (b) Convergence of reduced energy density $e$ towards the exact value $e_{Bethe}$ as a function of iteration number \cite{econv}.
}\label{fig:tdvp_vs_newopt}
\end{figure}

\Fig{fig:tdvp_vs_newopt} (a) (blue dots) illustrates the fast and robust convergence of the cMPS with the number of iterations 
of steepest descent, by showing the energy density $E \equiv \braket{H}$, particle density $\rho \equiv  \braket{\psi^\dagger\psi}$,
and reduced energy density $e$,
\begin{equation}
e \equiv \frac{E-\mu\rho}{\rho^3} = \braket{\frac{\partial_x\psi^{\dagger}\partial_x\psi}{2m}+g\psi^{\dagger}\psi^{\dagger}\psi\psi} / \braket{\psi^\dagger\psi}^3,
\end{equation}
for bond dimension $D=16$ and the choice of parameters $(\mu, g, m) = (-0.5, 1.0, 0.5)$. 
For comparison, we also show the same quantities when the cMPS is optimized instead by an Euclidean time
evolution using the TDVP algorithm (green crosses), starting from the same initial state and using values $d\tau=\alpha=0.01$
for TDVP and for the steepest descent optimization \cite{suppl}, respectively. These values for $d\tau$ are typically used in common TDVP calculations for cMPS \cite{draxler_private}.
\Fig{fig:tdvp_vs_newopt} (b) then shows the convergence of the energy $e$ to the exact value $e_{Bethe}$ obtained 
from the Bethe ansatz solution \cite{econv} as a function of iteration number, again for a steepest descent (blue dots) and
TDVP (green crosses) optimization. In this example, energy minimization 
converges towards the ground state roughly a hundred times faster than TDVP. The difference in performance is even bigger for larger 
bond dimension $D$, and/or when no lattice optimization is used to initialize the cMPS, in which case TDVP may even fail to 
converge.

\begin{figure}
  \includegraphics[width=1\columnwidth]{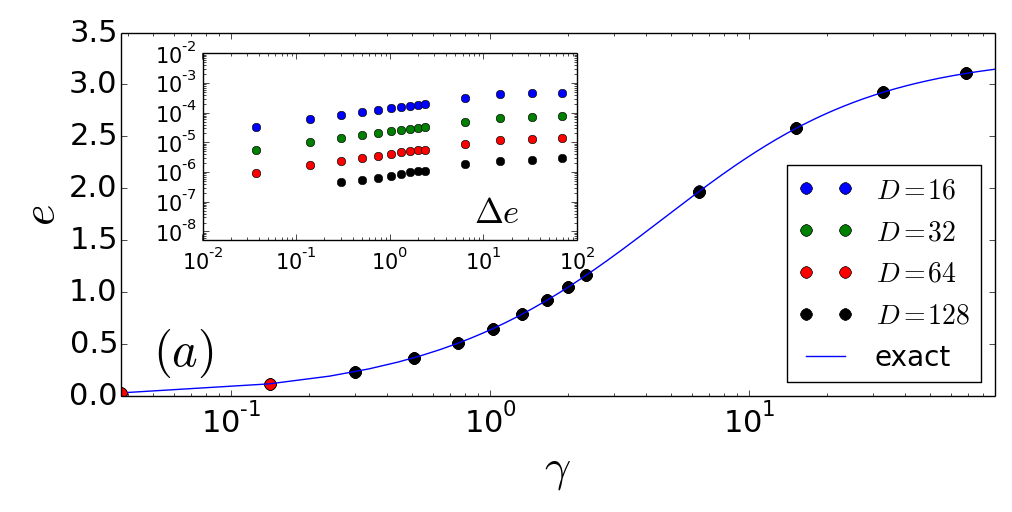}
  \includegraphics[width=0.49\columnwidth]{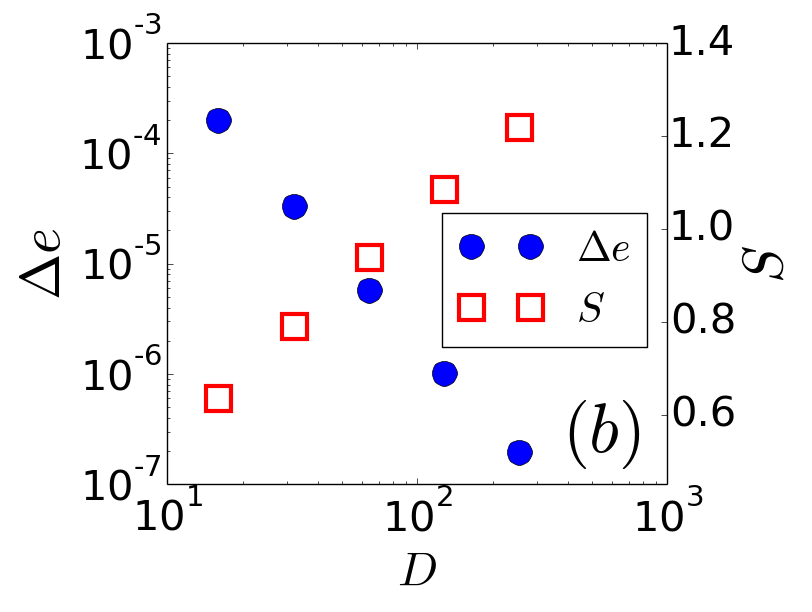}   
  \includegraphics[width=0.49\columnwidth]{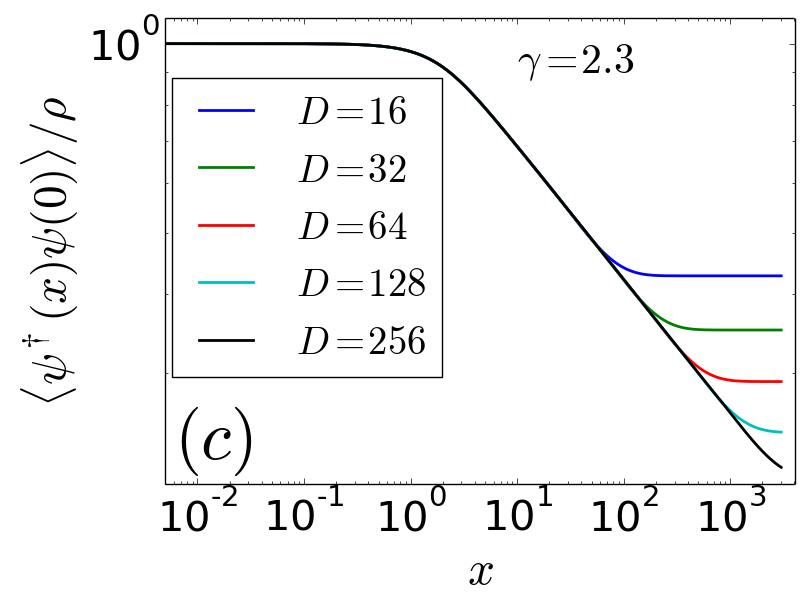}
\caption{
(a) Reduced energy density $e$ as a function of the dimensionless interaction strength $\gamma\equiv g/\rho$ and cMPS bond dimensions $D$. The solid line is the exact result from a Bethe ansatz calculation. Data points for different $D$ are on top of each other. The inset shows the error $\Delta e \equiv (e - e_{Bethe})/e_{Bethe}$.
(b) Relative error $\Delta e$ in the reduced energy density (filled circles) and bipartite entanglement $S$ (empty squares) of a left/right bipartition, as a function of the bond dimension.
(c) Superfluid correlation function, showing saturation to a constant at a finite correlation length $\xi$, 
  which diverges with growing $D$. 
} 
\label{fig:LLEs}
\end{figure}

\Fig{fig:LLEs}(a) illustrates the performance of the proposed energy minimization algorithm as a function of the bond dimension $D$.
For $D=16,32,64,128$, we computed the reduced energy density $e(\gamma)$ for several values of the dimensionless interaction 
strength $\gamma \equiv g/\rho$ in the range $[0.04,80]$ and observed a uniform pattern of convergence towards the exact 
$e_{Bethe}(\gamma)$. For reference, a $D=64$ optimization employing a non-linear conjugate gradient optimization \cite{suppl}
(stopped once the 
energy $E$ has converged to 9 digits) takes $\sim$6 minutes on a desktop computer \cite{mypc}, including  both the lattice 
initialization ($\sim$2 minutes) and the non-linear conjugate gradient optimization in the continuum ($\sim$4 minutes). This value of the bond dimension is 
the largest reported so far using TDVP \cite{draxler_particles_2013, stojevic_conformal_2015}. 

\Fig{fig:LLEs}(b)-(c) specializes to $\gamma\sim 2.3$, $e \sim 1.2$, and considers even larger values of the bond dimensions $D$, up to $256$, to reproduce well-understood finite-$D$ effects of the cMPS representation \cite{stojevic_conformal_2015, pirvu_matrix_2012, tagliacozzo_scaling_2008, pollmann_theory_2009}. \Fig{fig:LLEs}(b) shows the relative error $\Delta e$  in the reduced energy density
and the entanglement entropy $S \equiv - \sum_\alpha (\lambda_\alpha)^2 \log_2 (\lambda_\alpha)^2$ across a left/right bipartition, \Eq{eq:schmidt}. As expected, $\Delta e$ vanishes with $D$ as a power-law, $\Delta e \sim D^{p_1}$, whereas the entanglement entropy diverges logarithmically, $S \sim \log D$. \Fig{fig:LLEs}(c) shows the superfluid correlation function 
$\braket{\psi^{\dagger}(x)\psi(0)}/\rho$, which is seen to saturate to a finite value $|\braket{\psi}|^2/\rho$ at some distance $\xi$, another well-understood artefact of the (c)MPS representation at finite bond dimension $D$ \cite{stojevic_conformal_2015, pirvu_matrix_2012, tagliacozzo_scaling_2008, pollmann_theory_2009}. This artificial finite correlation length $\xi$ is seen to diverge with growing $D$ as a power-law, $\xi \sim D^{p_2}$.

\begin{figure}
  \includegraphics[width=0.49\columnwidth]{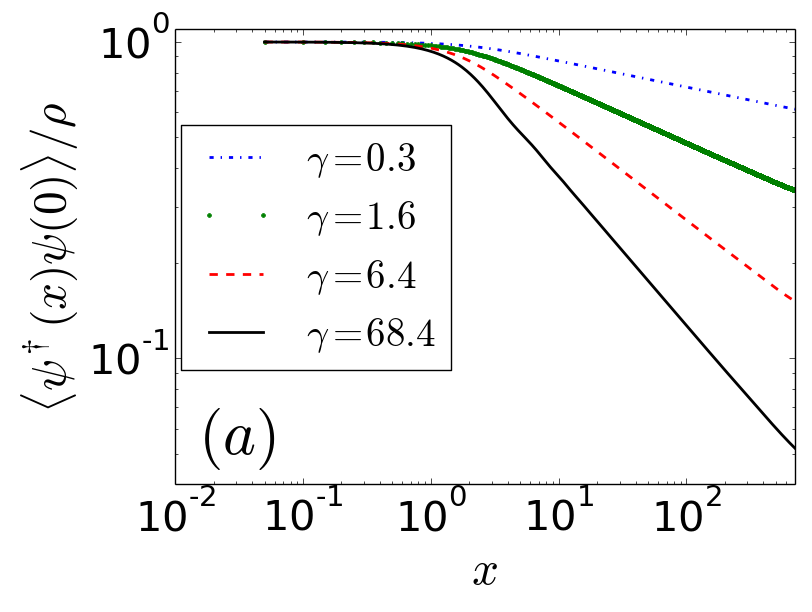}
  \includegraphics[width=0.49\columnwidth]{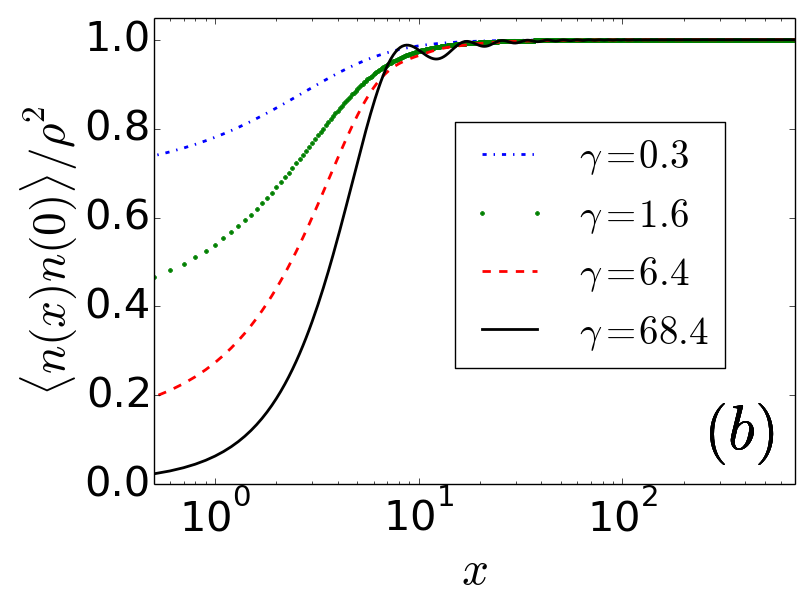}
  \caption{
    (a) Superfluid correlation function and (b) pair correlations as a function of the interaction strength $\gamma$, for $\mu=-0.5,D=128$.}\label{fig:corrs}
\end{figure}

Once we have established that the optimized cMPS is an accurate approximation to the ground state, we can move to exploring other properties of the model. \Fig{fig:corrs} 
shows the superfluid correlation function $\braket{\psi^{\dagger}(x)\psi(0)}/\rho$ and pair correlation function 
$\braket{n(x)n(0)}/\rho^2$, respectively, for $D=128$ and different values of the dimensionless interaction strength $\gamma$. 
With growing $\gamma$ we observe an increasingly rapid decay in the superfluid correlation function. 
The pair correlation function develops 
typical oscillations that are related to the fermionic nature of the ground state of the Tonks-Girardeau gas 
\cite{girardeau_relationship_1960} at $g=\infty$. 

We can also estimate both the central charge $c$ and the Luttinger parameter $K$, which can be used to uniquely identify the conformal 
field theory that characterizes the universal low energy / large distance features of the model. The central charge $c$ can be estimated from the slope of $S(D)$ (see Ref. \cite{stojevic_conformal_2015}). 
For $\gamma\approx 2.3$ we obtain a value of $c\approx 0.997$, to be compared with the exact value $c=1$. 
The Luttinger parameter $K$ \cite{cazalilla_bosonizing_2004,cazalilla_one_2011} is obtained from fitting $\log(\braket{\psi^{\dagger}(x)\psi(0)}/\rho)$ vs $\log(x)$ \cite{rincon_lieb-liniger_2015}, 
where we choose $x$ to lie in the region where $\braket{\psi^{\dagger}(x)\psi(0)}/\rho$ exhibits power-law decay. For $D=256$ and $\gamma\approx 2.3$ we obtain $K=2.362\pm 0.002$.
A value of $K=2.378$ was obtained in \cite{cazalilla_bosonizing_2004} from the weak-coupling approximation of the Bethe ansatz solution. 
The relative difference to our result is $\sim 0.7\%$.

\textit{Discussion.}--- The cMPS is a powerful variational ansatz for 
strongly interacting quantum field theories in 1+1 dimensions \cite{verstraete_continuous_2010}. In this Letter we have proposed a cMPS energy minimization algorithm with much better performance, in terms of convergence and the attainable bond dimension $D$, than previous optimization algorithms based on simulating an Euclidean time evolution. For benchmarking purposes, we have 
applied it to the exactly solvable Lieb Liniger model, but it performs equally well for a large variety of (non-exactly solvable) field theories \cite{suppl}.
We envisage that this algorithm will play a decisive role in unlocking the full potential of the cMPS representation for ground state studies in the continuum. 

Our algorithm works best by initializing the cMPS through an energy optimization on the lattice and by translating the resulting MPS from the lattice to the continuum through \Eq{eq:disccmps}. 
A natural question is then whether the continuum algorithm is needed at all. That is, perhaps --one may wonder-- an MPS algorithm working at finite lattice spacing $\epsilon$ can already provide a cMPS representation (through \Eq{eq:disccmps}) that can be made arbitrarily close to the one obtained with the continuum algorithm by decreasing $\epsilon$ sufficiently. The answer is that this is not possible: lattice algorithms necessarily become unstable as the lattice spacing $\epsilon$ is reduced. This can be understood from a simple scaling argument. In discretizing e.g. Hamiltonian $H$ of \Eq{eq:Ham} into a lattice, the non-relativistic kinetic term $\int \partial_x\psi^{\dagger}\partial_x\psi$ is seen to diverge with $\epsilon$ as $\sim 1/\epsilon^2$, while the rest of terms in the Hamiltonian have a milder scaling. For small $\epsilon$ this creates a large range of energy scales that lead to numerical instability. This effect is compounded with a second fact, revealed by \Eq{eq:disccmps}. For small $\epsilon$ the MPS matrix $A^0 = \mathbbm{1}+\epsilon Q$ is made of two pieces: a constant part $\mathbbm{1}$ made of $0$'s and $1$'s and the variational parameters $\epsilon Q$, which are of order $\epsilon$. Thus the first part $\mathbbm{1}$ \textit{shadows} the second one, in that the numerical precision on the variational parameters $Q$ is reduced by a factor $\epsilon$ when embedded in matrix $A^0$. The observant reader may then wonder if these problems could be prevented by just changing variables, to work instead with $Q = (A^0 - \mathbbm{1})/\epsilon$. This is indeed the case,  and also the essence of working with the cMPS representation directly, as we do in the proposed energy minimization algorithm. Notice that lattice MPS techniques can be succesfully applied to ground states \cite{dolfi_multigrid_2012, stoudenmire_one-dimensional_2012, wagner_guaranteed_2013, wagner_kohn-sham_2014, baker_one-dimensional_2015}
and real time evolution \cite{knap_quantum_2014, muth_dynamics_2010} of discretized field theories. However, these simulations are conducted at sufficiently large $\epsilon$ and are often plagued with finite $\eps$-scaling analysis, which is not necessary when working directly with a cMPS.

We have seen that the cMPS energy minimization algorithm drastically outperforms TDVP at the task of approximating the ground state (we emphasize that the TDVP remains an extremely useful tool e.g. 
to simulate real time evolution, for which no other method exists). This result did not come as a surprise: on the lattice, MPS energy minimization algorithms, including DMRG, have long been observed 
to converge to the ground state much faster than time evolution simulation algorithms \cite{mcculloch_infinite_2008}. 
We expect the new algorithm to also produce a significant speedup both for inhomogeneous Hamiltonians (where matrices $Q(x)$ and $R(x)$ depend on space \cite{verstraete_continuous_2010,haegeman_quantum_2015}), or for a theory of multiple fields $\psi_{\alpha}(x)$ \cite{haegeman_applying_2010, quijandria_continuous_2014, quijandria_continuous-matrix-product-state_2015, chung_matrix_2015,chung_multiple_2016,draxler_thesis_2016,haegeman_thesis_2011}), as we will discuss in future work.


\textit{Acknowledgements}
The authors thank D. Draxler, V. Zauner-Stauber, A. Milsted, J. Haegeman, F. Verstraete, and E. M. Stoudenmire for useful 
comments and discussions. The authors also acknowledge support by the Simons Foundation (Many Electron Collaboration). Computations were made 
on the supercomputer Mammouth parall\`ele II from University of Sherbrooke, managed by Calcul Qu\'ebec and 
Compute Canada. The operation of this supercomputer is funded by the Canada Foundation for Innovation (CFI), the 
minist\`ere de l'\'Economie, de la science et de l'innovation du Qu\'ebec (MESI) and the Fonds de recherche du Qu\'ebec - 
Nature et technologies (FRQ-NT). Some earlier computations were conducted at the supercomputer of the Center for Nanophase
Materials Sciences, which is a DOE Office of Science User Facility.
This research was supported in part by Perimeter Institute for Theoretical Physics. Research at Perimeter Institute is supported 
by the Government of Canada through Industry Canada and by the Province of Ontario through the Ministry of Economic Development \& 
Innovation.

\clearpage
\newpage

\pagebreak

\begin{center}
  \textbf{\large Supplementary Material}
\end{center}

\setcounter{equation}{0}
\setcounter{figure}{0}
\setcounter{table}{0}
\setcounter{page}{1}
\renewcommand{\theequation}{S\arabic{equation}}
\renewcommand{\thefigure}{S\arabic{figure}}
\renewcommand{\bibnumfmt}[1]{[S#1]}

\section{DMRG preconditioning}\label{app:dmrg}
As mentioned in the Main Text, our cMPS gradient optimization works best when using porperly prepared initial states. The purpose of
the following chapter is to introduce a novel method that employs DMRG energy minimization to obtain cMPS on a discretized system. 
These discrete states are then taken as initial states to a cMPS optimization in the continuum.
The main result here is that DMRG can be used to optimize an MPS with tensors of the form
\begin{eqnarray}\label{eq:disccmps2}
  \left( \begin{array}{c} A^{0} \\ A^{1} \end{array} \right) = 
  \left( \begin{array}{c} \mathbbm{1}+\Delta x Q \\ \sqrt{\Delta x} R \end{array} \right), 
\end{eqnarray}
in such a way that the cM
PS structure, i.e. the matrices $Q,R$, is always explicitly available. 
We use $\Delta x$ instead of $\eps$ to highlight that this is a lattice MPS. Note that even though we are 
treating soft core bosons here, we are restricting the local Hilbert space
dimension to $d=2$ (see below). 
We start by discretizing the Hamiltonian $H$ from 
Eq.(15) on a grid with spacing $\Delta x$.
With the replacement $\frac{1}{\sqrt{\Delta x}}c_i\equiv\psi(x_i)$, and using a first order discretization of the kinetic energy, this results in the following local 
Hamiltonian terms:
\begin{align}
  dx& \partial_x\psi^{\dagger}\partial_x\psi\approx\frac{1}{(\Delta x)^2}(c_{i+1}^{\dagger}-c_{i}^{\dagger})(c_{i+1}-c_{i})\label{eq:disckin}\\
  dx& \psi^{\dagger}(x)\psi^{\dagger}(x)\psi(x)\psi(x)\approx\frac{1}{\Delta x}c_{i}^{\dagger}c_{i+1}^{\dagger}c_{i}c_{i+1}\label{eq:discint}\\
  dx& \psi^{\dagger}(x)\psi(x)\approx c_i^{\dagger}c_i.\label{eq:discpot}
\end{align}
Note that we have replaced the local interaction term with a nearest neighbour interaction term. This replacement becomes exact in the limit $\Delta x\ra 0$.
By comparing the cMPS expressions for the local energy density 
\begin{align}
  \braket{h}=\frac{1}{2m}(l|[Q,R]\otimes[Q^*,R^*]|r)+\nonumber\\
  g(l|R^2\otimes R^{*2}|r)+\mu (l|R\otimes R^*|r)\nonumber
\end{align}
where $(l|$ and $|r)$ are the left and right reduced density matrices (see below),
with the ones obtained using the above discretization Eqs.(\ref{eq:disckin})-(\ref{eq:discpot}) and MPS tensors of the form
\Eq{eq:disccmps2}, it is easy to see that they can be made to coincide if in the numerical derivative \Eq{eq:disckin} one makes the replacement
\begin{align}\label{eq:modisckin}
\frac{1}{(\Delta x)^2}(c_{i+1}^{\dagger}-c_{i}^{\dagger})(c_{i+1}-c_{i})\to\nonumber\\
\frac{1}{(\Delta x)^2}(\mathcal{P}^0_ic_{i+1}^{\dagger}-c_{i}^{\dagger}\mathcal{P}^0_{i+1})(\mathcal{P}^0_{i}c_{i+1}-c_{i}\mathcal{P}^0_{i+1}),
\end{align}
where $\mathcal{P}^0_{i}$ is a projector onto the state with no particles at position $x_i$.
Thus, using \Eq{eq:discint} and \Eq{eq:discpot} and a modified kinetic energy \Eq{eq:modisckin} with MPS matrices 
of the form \Eq{eq:disccmps2} is almost identical to a discretized cMPS description as obtained in 
\cite{haegeman_quantum_2015}. The difference lies 
in the normalization: whereas MPS tensors \Eq{eq:disccmps2} are normalized according to a finite $\Delta x$, a true cMPS is normalized with respect to $\Delta x=0$.
The use of the projector $\mathcal{P}^0_i$ is crucial to the restriction of the local Hilbert space dimension to $d=2$.
The final Hamiltonian then assumes the form
\begin{align}
  H=\sum_{i\in \mathbbm{N}}-\frac{1}{2m(\Delta x)^2}(c_{i}^{\dagger}c_{i+1}+c_{i+1}^{\dagger}c_{i})+\frac{g}{\Delta x}c_i^{\dagger}c_{i+1}^{\dagger}c_ic_{i+1}\nonumber\\
  +\frac{1}{2m(\Delta x)^2}(\mathcal{P}^0_in_{i+1}+n_i\mathcal{P}^0_{i+1})+\mu_ic_i^{\dagger}c_i.\nonumber
\end{align}
For a given $\Delta x$, we then use standard iDMRG on a two site unit cell \cite{mcculloch_infinite_2008} to obtain the ground state. 
We note that our proposed gradient optimization for cMPS could, with minor modifications, also be applied to the discrete setting, and to homogeneous lattice
systems in general \cite{inprep}.
Once the iDMRG has been sufficiently converged, one can use 
the matrices $Q, R$ to initialize a new DMRG run for a finer discretization $\widetilde {\Delta x}<\Delta x$ by simply changing $\Delta x$ to $\widetilde {\Delta x}$
in \Eq{eq:disccmps2}, and renormalizing the MPS tensor. Using the 
terminology of Ref.\cite{dolfi_multigrid_2012}, we call this operation a prolongation
of the state. Even for $\Delta x$ as large as $\Delta x=0.1$ and for
the parameter values used in this Letter, this gives already good initial states for a cMPS optimization at $\Delta x=0$.
Note that since the matrices $Q,R$ depend on $\Delta x$, so does the quality of the prolonged initial state. 
This proposed fine graining procedure generalizes the approach presented in \cite{dolfi_multigrid_2012} to arbitrary prolongations.
Prior to prolonging the state to a finer discretization, it is vital to bring it into the canonical form 
\cite{vidal_efficient_2003} to remove any gauge jumps at the boundaries of the two-site unit cell. Note that in the diagonal gauge 
there is still a gauge freedom with diagonal, complex gauge matrices that has to be removed as well before prolongation. 
These gauge jumps originate from a unitary freedom in the SVD and QR decompositions, which are used heavily in lattice MPS 
optimizations.

\section{Determination of the Gradient for the Lieb Liniger model}\label{app:update}
In this section we elaborate on the determination of the gradient of the energy expectation value using cMPS methods (see also \cite{vanderstraeten_tangent} for details on lattice MPS gradients).
The goal is to optimize the matrices $Q_c, R_c$ to lower the energy 
\begin{equation}
  E(\lambda,Q_c,R_c)\equiv \frac{\mathcal{H}}{\mathcal{N}}\equiv \frac{\bra{\Psi}H\ket{\Psi}}{\braket{\Psi|\Psi}},\nonumber
\end{equation}
see Eq(13) in the main text.
Similar to DMRG \cite{schollwock_density-matrix_2011}, we calculate the gradient of this expression with respect to $Q_c^*,R_c^*$.
Taking the derivative $\delta (\mathcal{H}/\mathcal{N}) \equiv \frac{\partial}{\partial ({Q_c^*},R_c^*)} (\mathcal{H}/\mathcal{N})$ gives
\begin{align}
  &\delta (\mathcal{H}/\mathcal{N})=\frac{\delta \mathcal{H}}{\mathcal{N}}-\frac{\mathcal{H}}{\mathcal{N}^2}\delta \mathcal{N}.\nonumber
\end{align}
We normalize our state in such a way that $\mathcal{H}=0$ (see Eqs.(\ref{eq:renorml}) and (\ref{eq:renormr}) below), and thus the second term vanishes identically.
Since the energy depends non-linearly on $Q_c,R_c$, we have to apply the chain rule when calculating $\delta \mathcal{H}$.
However, due to translational invariance, this full gradient decomposes into a sum of $N$ identical terms, where each term
corresponds to a {\it local} gradient, and $N$ is the number of discretization points:
\begin{align}
  \delta \mathcal{H}=N (\delta \mathcal{H})_{local}.\nonumber
\end{align}
$(\delta \mathcal{H})_{local}$ is a {\it local} derivative, similar to DMRG, with respect to $Q_c^*, R_c^*$.
Note that for inhomogeneous systems, this local gradient would contain derivatives of the matrix $R$ \cite{haegeman_calculus_2013}.
Knowledge of this local gradient is thus sufficient to calculate the gradient of $\mathcal{H}$.
To obtain the local gradient $(\delta \mathcal{H})_{local}$, it is convenient to 
consider the state in the {\it center site form} 
\begin{align}
  \ket{\Psi}=
  \dots
  \left(
  \begin{array}{c}
    \mathbbm{1}+\eps Q_l\\
    \sqrt{\eps}R_l
  \end{array}
  \right)
  \left(
  \begin{array}{c}
    \lambda+\eps Q_c\\
    \sqrt{\eps}R_c
  \end{array}
  \right)
  \left(
  \begin{array}{c}
    \mathbbm{1}+\eps Q_r\\
    \sqrt{\eps}R_r
  \end{array}
  \right)
  \dots\; ,\nonumber
\end{align}
where $Q_c\equiv Q_l\lambda=\lambda Q_r,R_c\equiv R_l\lambda=\lambda R_r$. $\lambda$ is a diagonal matrix containing the Schmidt values.
$Q_{l/r}$ and $R_{l/r}$ are left/right normalized cMPS matrices, i.e.
\begin{align}
  (\mathbbm{1}|T_l\equiv (\mathbbm{1}|\left(Q_l\otimes \mathbbm{1}+\mathbbm{1}\otimes Q_l^* +R_l \otimes R_l^*\right)=0\nonumber\\
  T_r|\mathbbm{1})\equiv \left(Q_r\otimes \mathbbm{1}+\mathbbm{1}\otimes Q_r^*+R_r \otimes R_r^*\right)|\mathbbm{1})=0.\nonumber
\end{align}
Here we have defined the left/right cMPS transfer operator \cite{haegeman_calculus_2013}
\begin{align}\label{eq:cMPSTO_0}
  T_{l/r}\equiv Q_{l/r}\otimes \mathbbm{1}+\mathbbm{1}\otimes Q_{l/r}^* +R_{l/r} \otimes R_{l/r}^*.
\end{align}
The energy expectation $\mathcal{H}$ with respect to now {\it local} $Q_c,R_c$
(at the position where we want to take the local derivative)
can be evaluated to 
\begin{align}\label{eq:update0}
  &\braket{\Psi|H|\Psi}=\nonumber\\
  &\begin{array}{c}
    \lceil\\
    (H_l|\\
    \lfloor
  \end{array}
  \begin{array}{c}
    \left(
      \begin{array}{c}
        \lambda +\eps Q_c\\
        \sqrt{\eps} R_c
      \end{array}
    \right)\\
      |\\
    \left(
      \begin{array}{c}
        \mybar{\lambda^*} +\eps \mybar{Q_c^*}\\
        \sqrt{\eps} \mybar {R_c^*}
      \end{array}
    \right)
  \end{array}
  \Bigg]+
  \Bigg[
  \begin{array}{c}
    \left(
      \begin{array}{c}
        \lambda +\eps Q_c\\
        \sqrt{\eps} R_c
      \end{array}
    \right)\\
      |\\
    \left(
      \begin{array}{c}
        \mybar{\lambda^*} +\eps \mybar{Q_c^*}\\
        \sqrt{\eps} \mybar{R_c^*}
      \end{array}
    \right)
  \end{array}
  \begin{array}{c}
    \rceil\\
    |H_r)\\
    \rfloor
  \end{array}+\nonumber\\
  &\frac{\eps}{2m}\Bigg(\Big[
  \begin{array}{c}
    Q_lR_c-R_lQ_c\\
    \\
    \mybar Q_l^*\mybar{R_c^*}-\mybar R_l^*\mybar {Q_c^*}
  \end{array}
  \Big]
  +
  \Big[
  \begin{array}{c}
    Q_cR_r-R_cQ_r\\
    \\
    \mybar{Q_c^*} \mybar R_r^*-\mybar{R_c^*}\mybar Q_r^*
  \end{array}
  \Big]\Bigg)\nonumber\\
  &+\eps g\Bigg(\Big[
  \begin{array}{c}
    R_lR_c\\
    \\
    \mybar R_l^*\mybar{R_c^*}
  \end{array}
  \Big]+
  \Big[
  \begin{array}{c}
    R_cR_r\\
    \\
    \mybar{R_c^*}\mybar R_r^*
  \end{array}
  \Big]\Bigg)+
  \eps\mu
  \Big[
  \begin{array}{c}
    R_c\\
    \\
    \mybar{R_c^*}
  \end{array}
  \Big]
\end{align}
Symbols $[,],|,\lceil,\rceil,\lfloor,\rfloor$ represent tensor contractions. 
In DMRG, the matrices $(H_l|$ and $|H_r)$ in \Eq{eq:update0} are known as the left and right block Hamiltonians.
Below we describe how to calculate them in our context. We use a particular renormalization of these left and right
block Hamiltonians, such that 
\begin{align}
  \begin{array}{c}
    \lceil\\
    (H_l|\\
    \lfloor
  \end{array}
  \begin{array}{c}
    \lambda\\
    \\
    \\
    \mybar{\lambda^*}
  \end{array}\Bigg]=0\label{eq:renorml},
  \\
  \Bigg[
  \begin{array}{c}
    \lambda\\
    \\
    \\
    \mybar{\lambda^*}
  \end{array}
  \begin{array}{c}
    \rceil\\
    |H_r)\\
    \rfloor
  \end{array}=0.\label{eq:renormr}
\end{align}
Due to this renormalization, the expectation value $\mathcal{H}\equiv \braket{\Psi|H|\Psi}=0$.
The local derivative of $\mathcal{H}$ with respect to 
$Q_c^*,R_c^*$ is then given by
\begin{align}
  \eps \frac{\partial \mathcal{E}}{\partial Q_c^*}\equiv&  \frac{\partial \mathcal{H}}{\partial Q_c^*}=
  \eps\begin{array}{c}
    \lceil\\
    (H_l|\\
    \lfloor
  \end{array}
  \begin{array}{c}
      \begin{array}{c}
        \lambda\\
        \\
        \\
        \bullet \\
      \end{array}
  \end{array}
  \Bigg]+
  \eps
  \Bigg[
  \begin{array}{c}
      \begin{array}{c}
        \lambda\\
        \\
        \\
        \bullet \\
      \end{array}
  \end{array}
  \begin{array}{c}
    \rceil\\
    |H_r)\\
    \rfloor
  \end{array}+\label{eq:update_Q}\\
  &\frac{\eps}{2m}\Bigg(-\Big[
  \begin{array}{c}
    Q_lR_c-R_lQ_c\\
    \\
    R_l^*\quad\quad\quad\bullet
  \end{array}
  \Big]
  +
  \Big[
  \begin{array}{c}
    Q_cR_r-R_cQ_r\\
    \\
    \bullet\quad\quad\quad R_r^*
  \end{array}
  \Big]\Bigg)\nonumber\\
  \eps \frac{\partial \mathcal{E}}{\partial R_c^*}\equiv&  \frac{\partial \mathcal{H}}{\partial R_c^*}=
  \eps
  \begin{array}{c}
    \lceil\\
    (H_l|\\
    \lfloor
  \end{array}
  \begin{array}{c}
      \begin{array}{c}
        R_c\\
        \\
        \\
        \bullet \\
      \end{array}
  \end{array}
  \Bigg]+
  \eps
  \Bigg[
  \begin{array}{c}
      \begin{array}{c}
        R_c\\
        \\
        \\
        \bullet \\
      \end{array}
  \end{array}
  \begin{array}{c}
    \rceil\\
    |H_r)\\
    \rfloor
  \end{array}+\label{eq:update_R}\\
  &\frac{\eps}{2m}\Bigg(\Big[
  \begin{array}{c}
    Q_lR_c-R_lQ_c\\
    \\
    Q_l^*\quad\quad\quad\bullet
  \end{array}
  \Big]
  -
  \Big[
  \begin{array}{c}
    Q_cR_r-R_cQ_r\\
    \\
    \bullet\quad\quad\quad Q_r^*
  \end{array}
  \Big]\Bigg)\nonumber\\
  &+\eps g\Bigg(\Big[
  \begin{array}{c}
    R_lR_c\\
    \\
    \mybar R_l^*\bullet
  \end{array}
  \Big]+
  \Big[
  \begin{array}{c}
    R_cR_r\\
    \\
    \bullet\mybar R_r^*
  \end{array}
  \Big]\Bigg)+
  \eps\mu
  \Big[
  \begin{array}{c}
    R_c\\
    \\
    \bullet
  \end{array}
  \Big].\nonumber
\end{align}
$\bullet$ signs denote free matrix indices due to the removal of a matrix at this point 
in the tensor network. 
$\frac{\partial \mathcal{E}}{\partial Q_c^*}$ and $\frac{\partial \mathcal{E}}{\partial R_c^*}$ are then used to update $Q_c, R_c$:
\begin{align}
  Q_c\ra  Q= Q_c-\alpha \frac{\partial \mathcal{E}}{\partial Q_c^*}\nonumber\\
  R_c\ra  R= R_c-\alpha \frac{\partial \mathcal{E}}{\partial R_c^*}\nonumber.
\end{align}
At this point, the new state is of the form
\begin{align}
  \dots
  \lambda^{-1}
  \left(
  \begin{array}{c}
    \lambda+\eps Q\\
    \sqrt{\eps}R
  \end{array}
  \right)
  \lambda^{-1}
  \left(
  \begin{array}{c}
    \lambda+\eps Q\\
    \sqrt{\eps}R
  \end{array}
  \right)
  \lambda^{-1}
  \left(
  \begin{array}{c}
    \lambda+\eps Q\\
    \sqrt{\eps}R
  \end{array}
  \right)
  \dots\; .\nonumber
\end{align}
The matrices $\lambda^{-1}$ are now absorbed back into the cMPS tensors from e.g. the right side, i.e.
\begin{align}
  \left(
  \begin{array}{c}
    \lambda+\eps Q\\
    \sqrt{\eps}R
  \end{array}
  \right)
  \lambda^{-1}
\ra
  \left(
  \begin{array}{c}
    \mathbbm{1}+\eps Q\lambda^{-1}\\
    \sqrt{\eps}R\lambda^{-1}
  \end{array}
  \right)
  \equiv
  \left(
  \begin{array}{c}
    \mathbbm{1}+\eps \tilde Q\\
    \sqrt{\eps}\tilde R
  \end{array}
  \right)\nonumber.
\end{align}
$\tilde Q,\tilde R$ are then brought back into the central canonical gauge (see below), which produces a new triplet of matrices $(\lambda,Q_c,R_c)$ and completes one update step.
Convergence is measured by the norm of the gradient 
\begin{align}
  \nx\equiv\lVert \left(\partial E/\partial Q_c^*,\partial E/\partial R_c^*\right) \rVert=\nonumber\\
  \sqrt{\tr\big[\frac{\partial \mathcal{E}}{\partial Q_c^*}^{\dagger}\frac{\partial \mathcal{E}}{\partial Q_c^*}+\frac{\partial \mathcal{E}}{\partial R_c^*}^{\dagger} \frac{\partial \mathcal{E}}{\partial R_c^*}\big]}\nonumber.
\end{align}
This measure is similar to, but more stringent than, the one used in the TDVP,
in general $\lVert \dot x \rVert<\nx$ (see below for a definition of $\lVert \dot x\rVert$, the latter does not take $\frac{\partial \mathcal{E}}{\partial Q_c^*}$ into account).

Next we address the calculation of the matrices $(H_l|$ and $|H_r)$ in Eqs.(\ref{eq:update_Q}) and (\ref{eq:update_R}). In the context of DMRG, they are known as the left and right block Hamiltonians,
respectively. For an infinite, translational invariant system they are given by an infinite sum of the form
\begin{align}
  (H_l|=\eps(h_l|\big(\mathbbm{1}+\T_l+\T_l^2+\T_l^3+\dots\big)\label{eq:geosum1}\\
  |H_r)=\big(\mathbbm{1}+\T_r+\T_r^2+\T_r^3+\dots\big)|h_r)\eps\label{eq:geosum2}
\end{align}
where $\T_{l/r}$ is an infinitesimal MPS transfer operation, i.e. 
\begin{align}
  \T_{l/r}&=\mathbbm{1}\otimes \mathbbm{1}+\eps (Q_{l/r}\otimes \mathbbm{1}+\mathbbm{1}\otimes Q_{l/r}^*+R_{l/r}\otimes R_{l/r}^*)\nonumber\\
  &=\mathbbm{1}\otimes \mathbbm{1}+\eps T_{l/r},\nonumber
\end{align}
(see \Eq{eq:cMPSTO_0}) and
\begin{align}
  (h_l|=
  \frac{1}{2m}
  \Big[
  \begin{array}{c}
    Q_lR_l-R_lQ_l \\
    \\
    Q_l^*R_l^*-R_l^*Q_l^*
  \end{array}
  +g\Big[
  \begin{array}{c}
    R_l^2\\
    \\
    R_l^{*2}
  \end{array}
  +\mu
  \Big[
  \begin{array}{c}
    R_l\\
    \\
    R_l^*
  \end{array}\nonumber\\
  |h_r)=
  \begin{array}{c}
    Q_rR_r-R_rQ_r \\
    \\
    Q_r^*R_r^*-R_r^*Q_l^*
  \end{array}
  \Big]  \frac{1}{2m}+
  \begin{array}{c}
    R_r^2\\
    \\
    R_r^{*2}
  \end{array}
  \Big]g
  +
  \begin{array}{c}
    R_r\\
    \\
    R_r^*
  \end{array}\nonumber
  \Big]\mu
\end{align}
is related to the energy content of an infinitesimal interval $\eps$.
Since $\T_{l}$ has a left and right eigen-vector $(\mathbbm{1}|$ and $|\lambda^2)$ to eigenvalue 1 (and similarly $\T_r$  has eigenvectors
$|\mathbbm{1})$ and $(\lambda^2|$ ) the two sums in Eqs.(\ref{eq:geosum1}) and (\ref{eq:geosum2}) are divergent. These divergences can be regularized
by removing the subspace $|\lambda^2)(\mathbbm{1}|$ from $\T_{r}$ and  $|h_r)$, and $|\mathbbm{1})(\lambda^2|$ from $\T_l$ and $(h_l|$, i.e. by replacing
\begin{align}
  (h_l|\ra (h_l|_{\perp}\equiv(h_l|-(h_l|\lambda^2)(\mathbbm{1}|\nonumber\\
  \T_{l}\ra \T_{l}-\eps |\lambda^2)(\mathbbm{1}|\nonumber\\
  |h_r)\ra |h_r)_{\perp}\equiv|h_r)-|\mathbbm{1})(\lambda^2|h_r)\nonumber\\
  \T_{r}\ra \T_{r}-\eps |\mathbbm{1})(\lambda^2|\nonumber.
\end{align}
The geometric series can then be summed to 
\begin{align}\label{eq:HlHr}
  (H_l|=\eps(h_l|_{\perp}\frac{1}{1-\T_l+\eps|\lambda^2)(\mathbbm{1}|}=(h_l|_{\perp}\frac{1}{-T_l+|\lambda^2)(\mathbbm{1}|}\nonumber\\
  |H_r)=\frac{1}{1-\T_r+\eps|\mathbbm{1})(\lambda^2|}|h_r)_{\perp}\eps=\frac{1}{-T_r+|\mathbbm{1})(\lambda^2|}|h_r)_{\perp}
\end{align}
(note the cancellation of $\eps$ in \Eqs{eq:HlHr})
which is equivalent to 
\begin{align}
  \big[T_r-|\mathbbm{1})(\lambda^2|\big]|H_r)=-|h_r)_{\perp}\label{eq:Hl}\\
  (H_l|\big[T_l-|\lambda^2)(\mathbbm{1}|\big]=-(h_l|_{\perp}\label{eq:Hr}
\end{align}
with $T_{l/r}$ the left/right cMPS transfer operator \Eq{eq:cMPSTO}. We solve \Eqs{eq:Hl} and (\ref{eq:Hr}) using the lgmres routine provided by the scipy.sparse.linalg module \cite{pythonnote}.

Here is a summary of our proposed optimization scheme:

\begin{enumerate}
  \item Initialize a cMPS with matrices $Q,R$ and bring them into central canonical gauge. 
    Set a desired convergence $\varepsilon$.
  \item Calculate $(H_l|$ and $|H_r)$ according to \Eq{eq:Hl} and \Eq{eq:Hr}. \label{iterate}
  \item Calculate $\frac{\partial \mathcal{E}}{\partial Q_c^*}$ and $\frac{\partial \mathcal{E}}{\partial R_c^*}$ from Eqs. (\ref{eq:update_Q})
    and (\ref{eq:update_R}).
  \item Update:\label{enum:update}
    \begin{align}
      \tilde Q=(Q_c-\alpha \frac{\partial \mathcal{E}}{\partial Q_c^*})\lambda^{-1}\nonumber\\
      \tilde R=(R_c-\alpha \frac{\partial \mathcal{E}}{\partial R_c^*})\lambda^{-1}\nonumber
    \end{align}
  \item Regauge $\tilde Q,\tilde R$ into the central canonical form and measure 
    $\lVert \nabla E\rVert=\sqrt{\tr\big[\frac{\partial \mathcal{E}}{\partial Q_c^*}^{\dagger}\frac{\partial \mathcal{E}}{\partial Q_c^*}+\frac{\partial \mathcal{E}}{\partial R_c^*}^{\dagger}\frac{\partial \mathcal{E}}{\partial R_c^*}\big]}$.
    If $\lVert \nabla E\rVert<\varepsilon$ stop, otherwise go back to \ref{iterate}.
\end{enumerate}
During simulations we dynamically adapt the parameter $\alpha$, i.e. if we detect a large increase in $\lVert \nabla E\rVert$ we reject the update and redo the step with a smaller value of $\alpha$.

\section{Time Dependent Variational Principle for cMPS}\label{app:tdvp}
In the following we summarize the Time Dependent Variational Principle (TDVP) \cite{haegeman_time-dependent_2011,haegeman_unifying_2016,halimeh_enriching_2016} for homogeneous cMPS (see also the Appendix in \cite{rincon_lieb-liniger_2015} for more details).
To obtain the ground state of e.g. Eq.(15) in the Main Text, one employs Euclidean time evolution using the TDVP. The general strategy is the following:
given a cMPS $\ket{\Psi[Q(\tau),R(\tau)]}$ at time $\tau$, one finds an approximation $\ket{\Phi}\approx H\ket{\Psi[Q(\tau),R(\tau)]}$ that
can be used to evolve $\ket{\Psi[Q(\tau),R(\tau)]}$:
\begin{align}
  \ket{\Psi[Q(\tau+d\tau),R(\tau+d\tau)]}=
  \ket{\Psi[Q(\tau),R(\tau)]}-d\tau \ket{\Phi}\nonumber
\end{align}
The most general ansatz ansatz for $\ket{\Phi}$ is to superimpose local perturbations 
on the state $\ket{\Psi[Q(\tau),R(\tau)]}$:
\begin{align}\label{eq:tangentvec}
  &\quad\quad\quad\ket{\Phi(V,W)}\equiv \nonumber\\
  &\int_{-\infty}^{\infty}v_{-\infty}^{\dagger}dx\;U(-\infty,x)(V\otimes \mathbbm{1}+W\otimes \psi^{\dagger}(x))U(x,\infty)v_{\infty}|0\rangle,
\end{align}
where we $U(x,y)$ is
\begin{equation}
  U(x,y)=\mathcal{P}e^{\int_{x}^{y}dz\, Q\otimes \mathbbm{1}+R\otimes\psi^{\dagger}(z)}\nonumber.
\end{equation}
$v_{-\infty}$ and $v_{\infty}$ are irrelevant boundary vectors at $x=\pm\infty$.
The optimal $V_{opt}$ and $W_{opt}$ are found from minimizing the norm
\begin{align}\label{eq:normdist2}
  V_{opt}, W_{opt} = \text{argmin}_{\{V,W\}}\lVert\ket{\Phi}-H|\Psi\rangle\rVert^2,
\end{align}
with $\ket{\Phi(V,W)}$ defined in \Eq{eq:tangentvec}. This amounts to the calculation of tensor network expressions similar to Eqs.(\ref{eq:update_Q}) and (\ref{eq:update_R}).
The non-local nature of the perturbations in $\ket{\Phi(V,W)}$ leads to a complication in the minimization,
that can be overcome by resorting to a new parametrization $V_l,W$ (or $V_r,W$) such that the norm  $\lVert \ket{\Phi(V_l,W)}\rVert=\sqrt{\braket{\Phi(V_l^*,W^*)|\Phi(V_l,W)}}$ has
non-vanishing contributions only when $V_l,W$ and $V_l^*, W^*$ act at the same position in space. Two possible 
parametrizations are given by
\begin{align}\label{eq:tangauge}
  V_l=-\frac{1}{l}R^{\dagger}lW\nonumber\\
  V_r=-\frac{1}{r}WrR.
\end{align}
These are called the left or right tangent-space gauges, respectively. We note that this parametrization covers the full
tangent-space to the state $\ket{\Psi}$ \cite{haegeman_calculus_2013}.
Here, the only free parameter left is $W$. $l$ and $r$ are the left and right reduced density matrices
obtained as the left and right eigenvectors of the cMPS transfer operator 
\begin{align}\label{eq:cMPSTO}
  T=Q\otimes \mathbbm{1}+\mathbbm{1}\otimes Q^* +R \otimes R^*
\end{align}
to eigenvalue $\eta=0$, i.e. $(l|T=0\,(l|,\;\; T|r)=0\,|r)$ in braket notation. 
In the TDVP for cMPS, one further simplifies all expressions in the minimization by reparametrizing $W$ as 
\begin{align}
  W=\frac{1}{\sqrt{l}}Y\frac{1}{\sqrt{r}},\nonumber
\end{align}
where $Y$ is now the free parameter.
The convergence measure used in TDVP is given by the norm 
\begin{align}
  \lVert \dot x\rVert \equiv \lVert \ket{\Phi(V,W)}\rVert=\sqrt{\tr(W^{\dagger}lWr)}=\sqrt{\tr(Y^{\dagger}Y)}\nonumber.
\end{align}
For left normalized matrices $Q_l,R_l$, the update step is given by
\begin{align}
  Q_l\ra  Q=Q_l-d\tau V_l\nonumber\\
  R_l\ra  R=R_l-d\tau W\nonumber,
\end{align}
with $d\tau$ a small time step in imaginary time. $V_l$ and $W$ correspond to $\frac{\partial \mathcal{E}}{\partial Q_c^*}$
and $\frac{\partial \mathcal{E}}{\partial R_c^*}$ in the gradient optimization approach.

\begin{figure}
  \includegraphics[width=1\columnwidth]{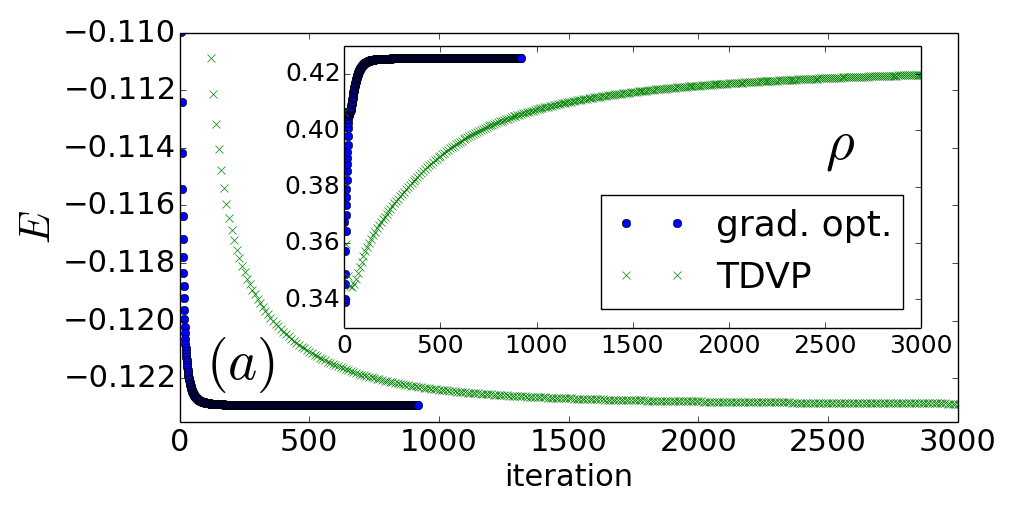}
  \includegraphics[width=1\columnwidth]{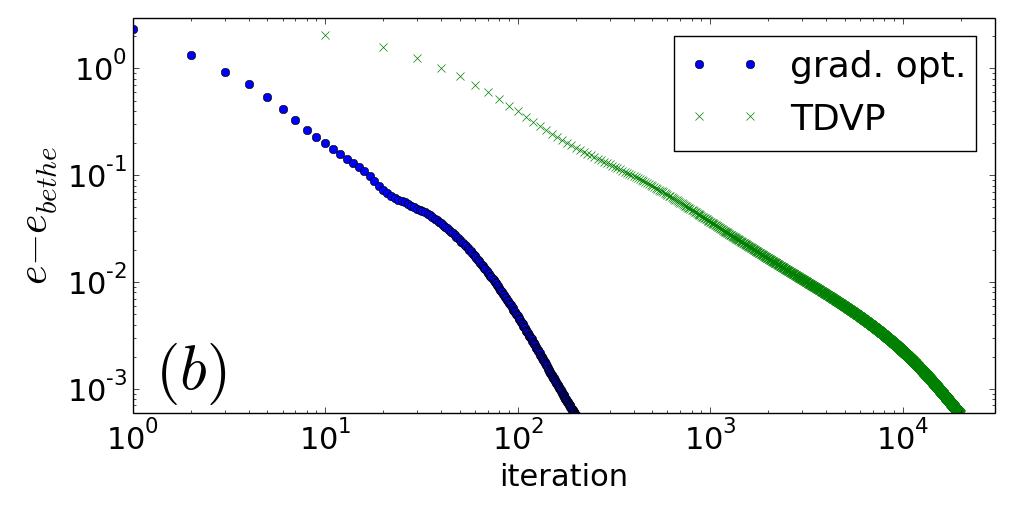}
  \caption{Comparison of cMPS gradient optimization with the TDVP, for $D=16$ and $\mu=-0.5,g=1.0,m=0.5$. We used a time
    step $\alpha=0.01$. (a) Energy density $E$ (main figure) and particle density
    $\rho$ (inset) as a function of iteration number. (b) Convergence of $e$ to the exact value from Bethe ansatz
    as a function of iteration number.}\label{fig:tdvp_vs_newopt_old}
\end{figure} 
\begin{figure}
  \includegraphics[width=0.85\columnwidth]{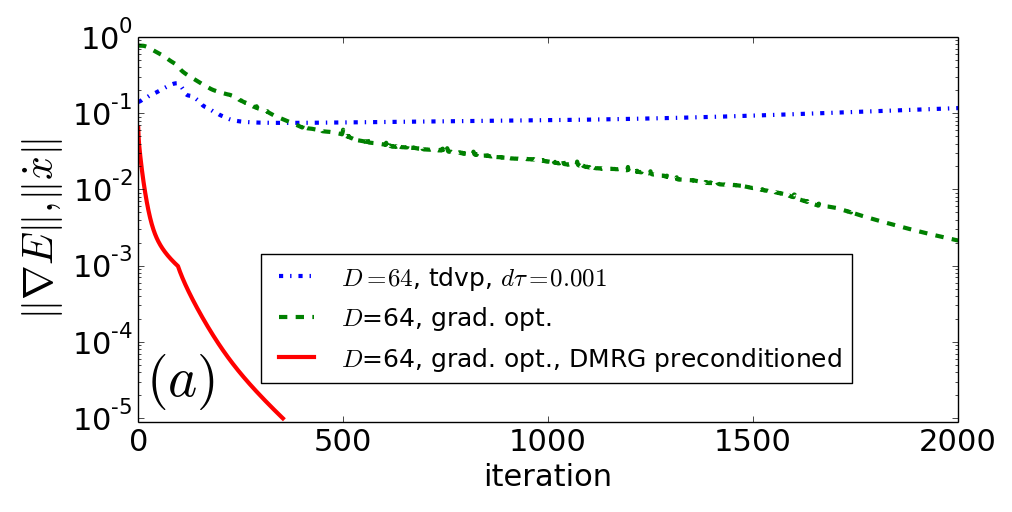}
  \includegraphics[width=0.85\columnwidth]{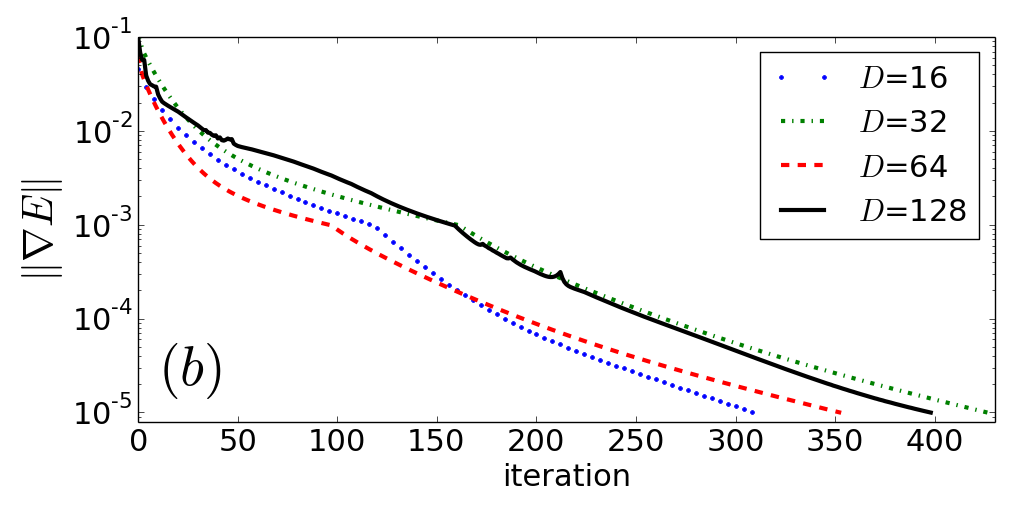}
  \caption{(a) Comparison of convergence of regular TDVP (blue dash-dotted) vs. cMPS {\it steepest descent} gradient optimization
    without (green dashed) and with (red solid) DMRG-preconditioning, for $D=64$ and $\mu=-0.5, g=1.0, m=0.5$. 
    The TDVP would take several $\mathcal{O}(10^4)$ steps to converge. 
    The gradient optimization without DMRG-preconditioning and dynamically adapted $\alpha$ took roughly 2500 iterations (4h runtime on a desktop PC \cite{mypc}). 
    With preconditioning it finished after 354 iterations 
    (which took $\approx 1043$ second or 17.4 min). 
    The DMRG preconditioner was run for $\Delta x=0.5,0.1$, and took roughly 140 seconds to converge \cite{mypc}
    We used $\alpha=0.01$ for $\nx>10^{-3}$ and $\alpha=0.03$ for $\nx\leq 10^{-3}$. 
    (b) $\nx$ of {\it steepest descent} gradient optimization as a function of iteration number for different bond dimensions $D$, with DMRG-preconditioning (total run time (including DMRG preconditioning) 
    for $D=128$ was 2.6h on a desktop PC \cite{mypc}).}\label{fig:DMRGprecond}
\end{figure} 
\begin{figure}
  \includegraphics[width=0.85\columnwidth]{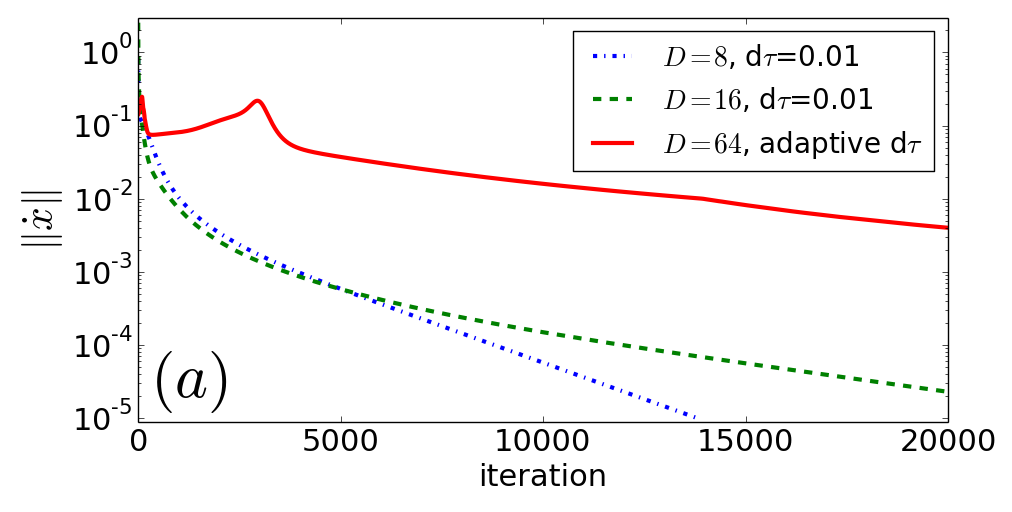}
  \includegraphics[width=0.85\columnwidth]{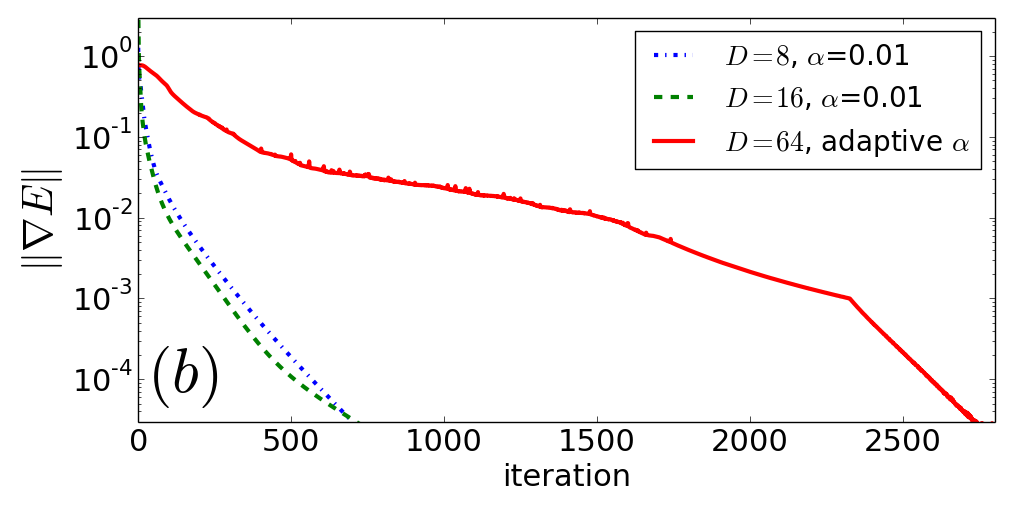}
  \caption{(a) TDVP ground state simulation: $\lVert \dot x\rVert$ as a function of iteration number for different bond dimensions $D$, and $\mu=-0.5, g=1.0, m=0.5$. For $D=64$ we used an 
    adaptive $d\tau$ scheme. (b) cMPS {\it steepest descent} gradient optimization: $\nx$ as a function of iteration number for different bond dimensions $D$, and $\mu=-0.5, g=1.0, m=0.5$. For $D=64$ we used
  the same adaptive $\alpha$ scheme as in (a).}\label{fig:tdvp_vs_newopt_diffD}
\end{figure} 
\begin{figure}
  \includegraphics[width=0.85\columnwidth]{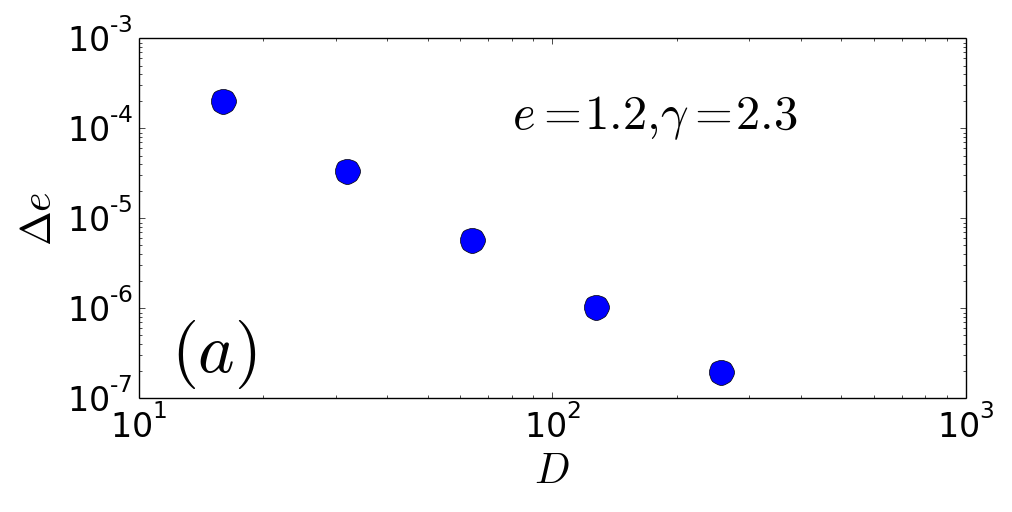}
  \includegraphics[width=0.85\columnwidth]{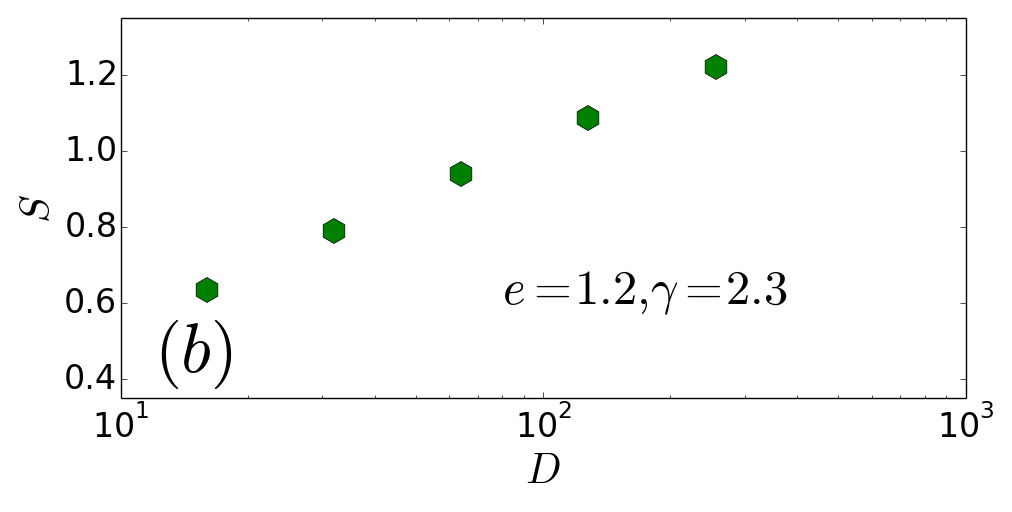}
  \includegraphics[width=0.85\columnwidth]{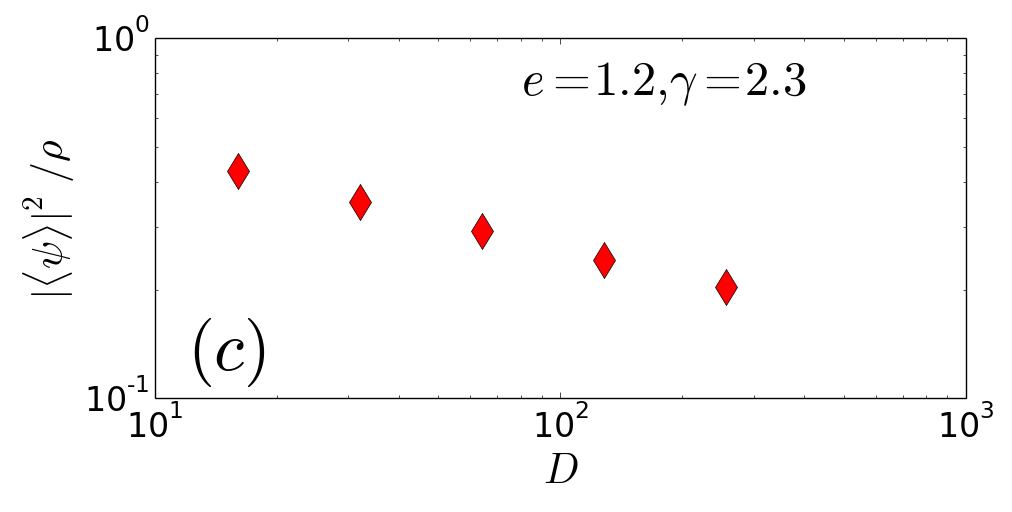}
  \caption{(a) Relative error of the dimensionless groundstate energy, (b) bipartite entanglement entropy $S \equiv - \sum_\alpha (\lambda_\alpha)^2 \log_2 (\lambda_\alpha)^2$
     and (c) order parameter $|\braket{\psi}|^2/\rho$ vs. bond dimension $D$, for $D=16,32,64,128,256$, 
     and $\mu=-0.5, \gamma\equiv g/\rho\approx 2.3\; (g=1.0)$.}\label{fig:Dscaling}
\end{figure}

Our gradient calculation includes one additional computational step as compared to the TDVP. 
There, depending on the choice of tangent-gauge,
the calculation of one of the two expressions containing $(H_l|$ or $|H_r)$ 
in Eqs.(\ref{eq:update_Q}) and (\ref{eq:update_R}) can be omitted. In our case we need 
to calculate both of them. This adds however only a small overhead in computational time.

\section{Obtaining the central canonical form}
An important ingredient to our gradient optimization is the {\it central canonical form}, introduced in the Main Text.
Here we describe how to obtain the central canonical form for cMPS. The procedure is 
a straight forward adaption of the corresponding lattice algorithm.
Starting from unnormalized matrices $Q,R$, one first calculates the left and right 
eigenvectors $(l|,|r)$ ($l,r$ in matrix notation) of the transfer operator $T$ \Eq{eq:cMPSTO} to the largest real eigenvalue 
$\eta$,
\begin{align}
  (l|T=\eta (l|\nonumber\\
  T|r)=\eta |r)\nonumber,
\end{align}
where $\eta\in \mathbbm{R}$. Next, the cMPS is normalized by
\begin{align}
  Q\ra Q-\frac{\eta}{2}\mathbbm{1}\nonumber.
\end{align}
This shifts the eigenvalue of $(l|$ and $|r)$ to $\eta'=0$.
Using $l,r$, compute $X=\sqrt{r},Y=\sqrt{l}$ and use a singular value decomposition to obtain
\begin{align}
  U\lambda D=YX,\nonumber
\end{align}
where $\lambda$ contains the Schmidt values.
Normalize $\lambda$ by
\begin{align}
  \lambda\ra \frac{\lambda}{\sqrt{\tr(\lambda\lambda^{\dagger})}}.\nonumber
\end{align}

The left or right normalized cMPS matrices $Q_l,R_l$ or $Q_r,R_r$ are then obtained from
\begin{align}
  Q_l=G_lQG_l^{-1},\nonumber\\
  R_l=G_lRG_l^{-1}\nonumber\\
  Q_r=G_rQG_r^{-1}\nonumber\\
  R_r=G_rRG_r^{-1}\nonumber
\end{align}
with
\begin{align}\label{eq:Gs}
  G_l=\lambda DX^{-1}\nonumber\\
  G_l^{-1}=Y^{-1}U\nonumber\\
  G_r=DX^{-1}\nonumber\\
  G_r^{-1}=Y^{-1}U\lambda.
\end{align}
Note that $Q_l\lambda=\lambda Q_r, R_l\lambda=\lambda R_r$.

\section{Non-linear conjugate gradient}
The non-linear conjugate gradient approach is an improvement of the steepest descent method that reuses gradient information from previous iterations, see e.g. \cite{nlcg}, and 
has recently been proposed as an improvement of the TDVP \cite{milsted_matrix_2013}.
At iteration $n$, the new search direction $\vec v_n$ is given by a linear combination of the current 
gradient $\nabla E_n$ and the previous search direction $\vec v_{n-1}$ from step $n-1$:
\begin{align}
  \vec v_n=\nabla E_n +\beta_n \vec v_{n-1}.
\end{align}
The parameter $\beta_n$ is calculated from a semi-empirical formula. In our case, we use 
the Fletcher-Reeves form
\begin{align}
  \beta_n=\frac{\lVert \nabla E_n\rVert^2}{\lVert \nabla E_{n-1}\rVert^2}.
\end{align}
In a regular implementation, one uses a line search method to find the minimum in the direction
$\vec v_n$ of $E$, which is then taken to be the starting point for the next iteration.
We omit this line search in our implementation.
The procedure is initialized with $\vec v_{-1}=0$. In our case, there is a small caveat due
to a gauge change of the state between iterations $n-1$ and $n$ when regauging the 
state into the central canonical form. Due to this gauge change, the current gradient and 
the previous search direction can only be added after the latter has been transformed 
into the gauge of the current state \cite{ashprivate}.
This is achieved using the matrices $G_l$ and $G_r$ from \Eq{eq:Gs}. 
Consider an update for the state $(\lambda^{[n]},Q_c^{[n]}, R_c^{[n]})$ at iteration $n$.
For the case of a left-sided update where the state is connected back by multiplication of 
$[\lambda^{[n]}]^{-1}$
from the right (see \ref{enum:update}. in the summary of our proposed algorithm),
the new search direction is given by
\begin{align}
  \vec v_n=
  \left(
  \begin{array}{c}
    \frac{\partial \mathcal{E}}{\partial Q_c^*}^{[n]}\\
    \frac{\partial \mathcal{E}}{\partial R_c^*}^{[n]}
  \end{array}\nonumber
  \right)[\lambda^{[n]}]^{-1}
  +
  \beta_n G_l^{[n]} \vec v_{n-1}[G_l^{[n]}]^{-1}
\end{align}
where $G_l^{[n]}$ is the gauge matrix obtained from gauging the state from the previous iteration
$n-1$ into the central canonical form at iteration $n$ (see \Eq{eq:Gs})
i.e.
\begin{align}
  Q_l^{[n]}=G_l^{[n]}\tilde Q^{[n-1]}[G_l^{[n]}]^{-1}\nonumber\\
  R_l^{[n]}=G_l^{[n]}\tilde R^{[n-1]}[G_l^{[n]}]^{-1}\nonumber.
\end{align}
The matrices $Q_c^{[n]},R_c^{[n]}$ are then updated according to 
\begin{align}
  \left(
  \begin{array}{c}
    \tilde Q^{[n]}\\
    \tilde R^{[n]}
  \end{array}\nonumber
  \right)
  =
  \left(
  \begin{array}{c}
    Q_c^{[n]}\\
    R_c^{[n]}
  \end{array}\nonumber
  \right)[\lambda^{[n]}]^{-1}-\alpha\vec v_n.
\end{align}
This algorithm only works reliably for $\nx$ below a certain threshold which
we empirically find to be $\nx<10^{-2}$. Further more, the Fletcher-Reeves update
is prone to jamming, i.e. stagnation of convergence rates. One way to overcome
this is by resetting the method after a certain number of $N_{reset}$ steps by setting $\beta_n=0$, 
thereby discarding all previous information after $N_{reset}$ iterations. We found
$N_{reset}=8-10$ to work best in our simulations.

\section{Further results}\label{app:further}
\subsection{Lieb Liniger Model}
In this section we show further comparisons of the cMPS steepest descent gradient optimization with the TDVP. \Fig{fig:tdvp_vs_newopt_old} shows results for convergence of $E$ and $\rho$ with
iteration number similar to Fig.(1) the main text, but with a random common initial state as opposed to one that 
has been preconditioned with DMRG. The results are in qualitative agreement with the ones in the Main Text.

In \Fig{fig:DMRGprecond} (a) we compare the convergence of regular TDVP (blue) without DMRG-preconditioning, the cMPS steepest 
descent gradient optimization scheme without DMRG-preconditioning (green) and 
cMPS steepest descent gradient optimization scheme with DMRG-preconditioning (red). In all simulations 
we used an adaptive $\alpha$ scheme to keep the simulation stable. We ran our DMRG preconditioner for a lattice discretization $\Delta x=0.5$ and 0.1 for 400 and 2000 DMRG steps,
respectively (the total run time of the two DMRG runs was roughly 140 seconds \cite{mypc}). 
We then prolonged the resulting state to $\Delta x=0.0$ and took it as initial state for a cMPS optimization.
We stopped the cMPS optimization once $\nx<5\times 10^{-5}$.
The simulation with DMRG-preconditioning converged within 354 iterations (17.4 min \cite{mypc}), as compared to 2500 iterations without preconditioning, and several ten thousand iterations for TDVP without
preconditioning. \Fig{fig:DMRGprecond} (b) shows $\nx$ for the cMPS steepest descent gradient optimization with DMRG-preconditioning for different values of $D$. The total runtime for the largest bond dimension $D=128$ 
was roughly 2.6h on a desktop PC \cite{mypc}. The plot demonstrates that the number of steps to converge a cMPS ground state does not depend on the bond dimension $D$ for our new scheme.

In \Fig{fig:tdvp_vs_newopt_diffD} (a) we show convergence of TDVP for different bond dimensions $D=8,16$ and 64, \Fig{fig:tdvp_vs_newopt_diffD} (b) shows
results for the cMPS steepest descent gradient optimization for the same cases. For $D=8,16$, we used a single time step $\alpha=0.01$ throughout the simulation, whereas for $D=64$ we 
resorted to an adaptive scheme. The computational gain of our proposed scheme clearly increases with increasing $D$. 
We note that the total reachable accuracy of the ground state in both the TDVP and our proposed scheme depends on the accuracy of the eigensolvers used to 
obtain $|r),(l|,|H_r)$ and $(H_l|$ \cite{pythonnote}. By using a higher accuracy in these solvers, the accuracy can be increased at the
cost of a longer run time per iteration step.

In \Fig{fig:Dscaling} (a) to (c) we plot the relative error of the reduced ground state energy, the entanglement entropy $S$ and the order parameter $|\braket{\psi}|^2/\rho$ 
for different values of $D=16,32,64,128,256$ and $m=0.5,\mu=-0.5,\gamma\equiv g/\rho\approx 2.3\; (g=1.0)$.

\subsection{Non-integrable models}

\begin{figure}
  \includegraphics[width=1\columnwidth]{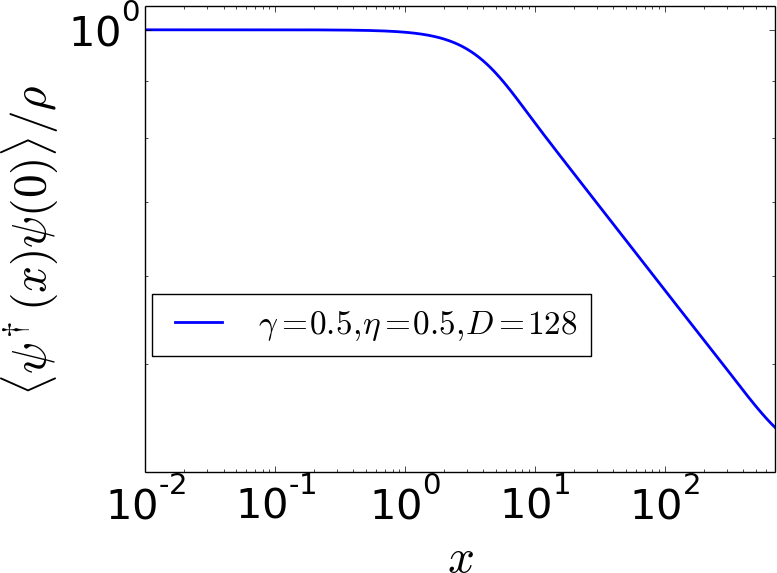}
  \includegraphics[width=1\columnwidth]{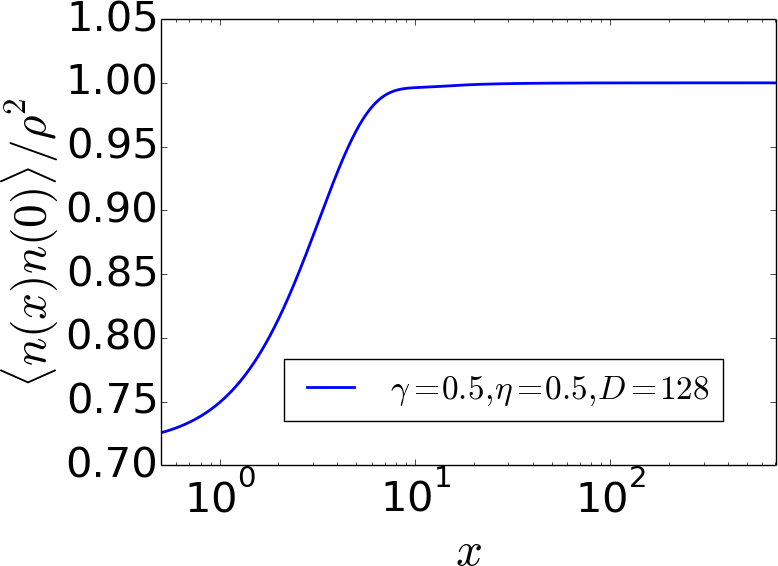}
  \caption{Structure factor (top) and pair correlation function (bottom) for 
    the ground state of the extended Lieb Liniger model, with $g_2=0.25$ 
    ($\gamma\equiv g_2/\braket{n}\approx 0.5),m=0.5, \mu=-0.5, \eta=0.5$ and 
    $D=128$.
}\label{fig:fig1}
\end{figure}

\begin{figure}
  \includegraphics[width=1\columnwidth]{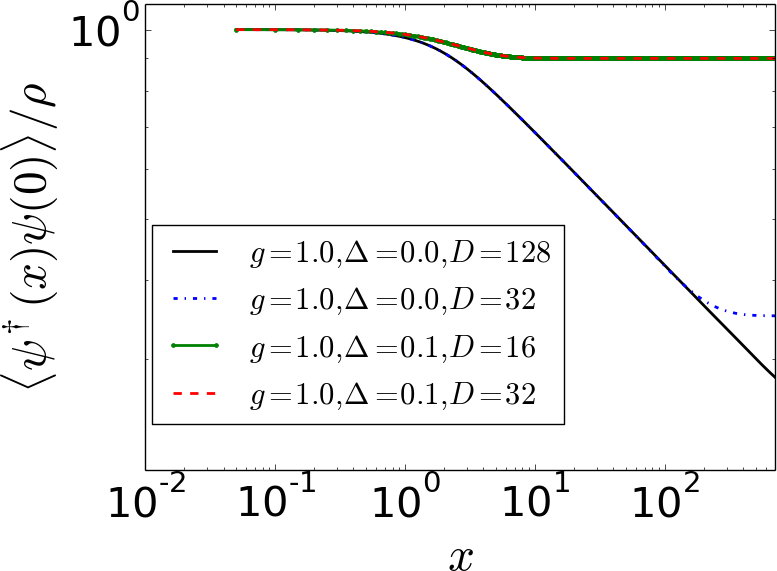}
  \includegraphics[width=1\columnwidth]{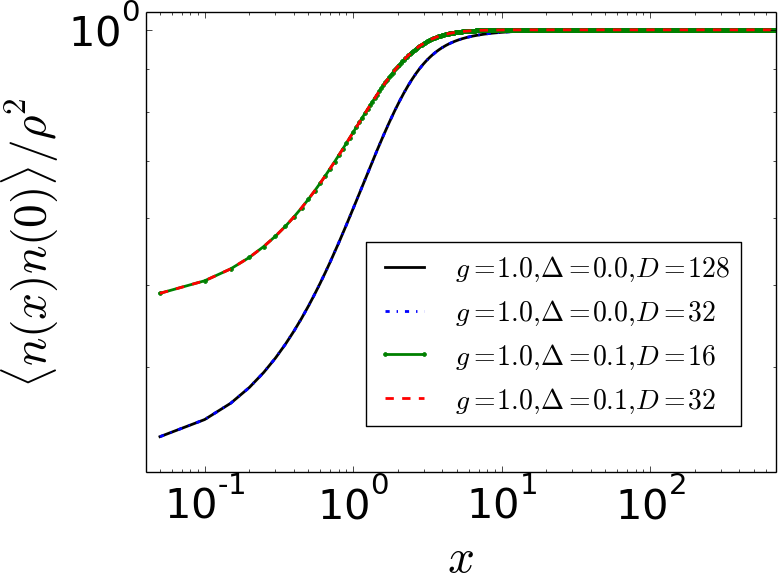}
  \caption{Structure factor (top) and pair correlation function (bottom) for 
    the ground state of a Lieb Liniger model with pairing interaction, 
    for $g=1,m=0.5,\mu=-0.5, \Delta=0.1$. Green dash-dotted  and red dashed 
    lines are results for $D=16$ and $D=32$, respectively. Increasing the bond
    dimension from $D=16$ to $D=32$ gives no significant change in the results.
    Thus, the saturation of the structure factor at a finite value reflects 
    the fact that the system is gapped. In contrast, for $\Delta=0.0$
    (blue dotted for $D=32$ and black lines for $D=128$) 
    the results are very sensitive to an increase in $D$. There, the saturation
    is an artifact of the finite bond dimension of the cMPS.
}\label{fig:fig2}
\end{figure} 

In this section we present additional results obtained with the gradient optimization algorithm. In particular, we obtain ground-state correlation functions for the following 
two Hamiltonians:\\
\textit{Extended Lieb Liniger model:}
In this model, the bosons are interacting with each other via a longer-range, exponentially suppressed interaction. The Hamiltonian takes on the form
\begin{align}
  H_1 &= \int dx\, \Big(\frac{1}{2m} \partial_x\psi^\dagger(x) \partial_x\psi(x) + \mu   \psi^\dagger(x) \psi(x)\nonumber\\
  &+ g_2\int_{y=x}^{\infty} dx\;dy\; e^{-\eta(y-x)}\,\psi^\dagger(x) \psi^\dagger(y) \psi(y) \psi(x)\Big),
\label{eq:HamELL}
\end{align}
with $\eta>0$ the range of the interaction, and $g_2$ the interaction strength. $\mu$ and $m$ are
chemical potential and mass, respectively.

\textit{Interacting bosons with a U(1) breaking pairing term:}
The second Hamiltonian is a Lieb-Liniger type Hamiltonian with an additional pairing term:
\begin{align}
  H_2 &= \int dx\, \Big(\frac{1}{2m} \partial_x\psi^\dagger(x) \partial_x\psi(x) + \mu   \psi^\dagger(x) \psi(x)\nonumber\\
  &+ g \,\psi^\dagger(x) \psi^\dagger(x) \psi(x) \psi(x)+\Delta\psi^{\dagger}(x)\psi^{\dagger}(x)\psi(x)\psi(x)\Big),
\label{eq:HamP}
\end{align}
where $\Delta$ is a pairing term which brakes the U(1) symmetry of the Lieb Liniger model, and $g,\mu$ and $m$ are interaction strength,
chemical potential and mass, respectively.
Both Hamiltonians are no longer integrable, and can thus not be solved exactly by Bethe's Ansatz. In particular, \Eq{eq:HamP} has a gapped 
ground state for $\Delta\neq 0$. The additional contributions to the gradient due to the pairing and exponential interaction terms are calculated similar to
e.g. \cite{draxler_particles_2013,rincon_lieb-liniger_2015}.

The results are shown in \Fig{fig:fig1} and \Fig{fig:fig2}. \Fig{fig:fig1} shows
the structure factor and pair correlation function for the extended Lieb 
Liniger model (parameters are in the caption). The structure factor shows 
algebraic decay over several decades. The model is in this case critical, and
can be described by Luttinger Liquid theory.

\Fig{fig:fig2} shows results for a Lieb Liniger type model with an additional 
pairing term $\Delta=0.1$. For $\Delta\neq 0$, the model is gapped. A bond 
dimension of $D=16$ is in this case already sufficient to obtain an accurate 
wave-function. This can be seen from the top panel in \Fig{fig:fig2}, where
we show results for the structure factor for $D=16,32$ 
(green dotted, red dashed lines). The two curves are practically on top of each 
other. The saturation at a finite value is thus of physical origin.
In contrast, for $\Delta=0.0$, the model is critical, and a saturation
of the structure factor in this case is an artifact due to the finite cMPS
bond dimension. This is illustrated by the blue dash-dotted and black curves 
in the top panel of \Fig{fig:fig2}, which show results for $\Delta=0$ and 
$D=32$ and $128$, respectively. There, an increase in $D$ from 32 to 128
shows a significant change in the results. The bottom panel in \Fig{fig:fig2} 
shows the pair correlation function for the same parameters.

\begin{thebibliography}{64}
\bibitem[{mor()}]{moreA}
\bibinfo{howpublished}{In Eq. (5) we neglect other possible matrices $A^{p}
  \equiv (\eps^{p/2}/p!) R^p$ for $p>1$ that are not relevant to the current
  discussion.}
\bibitem[{Inf()}]{InfiniteChain}
\bibinfo{howpublished}{In the limit of an infinite periodic lattice,
  $L\rightarrow \infty$, we can use the Schmidt decomposition of $\ket{\Psi}$
  as in \Eq{eq:schmidt}, which assumes a lattice with open boundary conditions.
  Further, in the case of a translation invariant system on an infinite chain,
  all left/right partitions are equivalent.}
\bibitem[{eco()}]{econv}
\bibinfo{howpublished}{As reference value we take
  $e_{Bethe}(\gamma_{\infty}=g/\rho_{\infty})$, where $\rho_{\infty}$ is the
  converged particle density at large iteration number of the TDVP or the
  gradient optimization, respectively.}
\bibitem[{pyt()}]{pythonnote}
\bibinfo{howpublished}{Note that special care has to be taken when using the
  scipy.linalg.eigs function to calculate left and right eigenvectors of the
  transfer operator, due to erroneous behaviour of the routine.}
\end{thebibliography}
\end{document}